\documentclass[AMA,STIX1COL]{WileyNJD-v2}
\usepackage[T1]{fontenc}
\usepackage[latin9]{inputenc}
\usepackage{color}
\usepackage{float}
\usepackage{textcomp}
\usepackage{amsmath}
\usepackage{graphicx}

\makeatletter

\floatstyle{ruled}
\newfloat{algorithm}{tbp}{loa}
\providecommand{\algorithmname}{Algorithm}
\floatname{algorithm}{\protect\algorithmname}

\articletype{Full Paper}

\received{}
\revised{}
\accepted{}

\DeclareMathOperator*{\argmin}{arg\,min}

\@ifundefined{showcaptionsetup}{}{%
 \PassOptionsToPackage{caption=false}{subfig}}
\usepackage{subfig}
\makeatother

\begin{document}
\title{Fast Data-Driven Learning of MRI Sampling Pattern for Large Scale Problems}

\author[1]{Marcelo V. W. Zibetti*}
\author[1]{Gabor T. Herman}
\author[3]{Ravinder R. Regatte}

\authormark{Zibetti \textsc{et al}}

\address[1]{\orgdiv{Center for Biomedical Imaging, Department of Radiology}, \orgname{New York University School of Medicine}, \orgaddress{New York \state{NY}, \country{USA}}}

\corres{*Marcelo V. W. Zibetti,  New York University School of Medicine, Department of Radiology, 660 First Avenue, New York, NY, 10016, USA. \email{Marcelo.WustZibetti@nyulangone.org}}

\fundingInfo{This study is supported by NIH grants R21-AR075259-01A1, R01-AR076328, R01-AR067156, and R01-AR068966, and was performed under the rubric of the Center of Advanced Imaging Innovation and Research (CAIR), an NIBIB Biomedical Technology Resource Center (NIH P41-EB017183).}

\abstract[]{\textbf{Purpose:} A fast data-driven optimization approach, named bias-accelerated subset selection (BASS), is proposed for learning efficacious sampling patterns (SPs) with the purpose of reducing scan time in large-dimensional parallel MRI. \\\textbf{Methods:} BASS is applicable when Cartesian fully-sampled k-space data of specific anatomy is available for training and the reconstruction method is specified, learning which k-space points are more relevant for the specific anatomy and reconstruction in recovering the non-sampled points. BASS was tested with four reconstruction methods for parallel MRI based on low-rankness and sparsity that allow a free choice of the SP. Two datasets were tested, one of the brain images for high-resolution imaging and another of knee images for quantitative mapping of the cartilage. \\\textbf{Results:} BASS, with its low computational cost and fast convergence, obtained SPs 100 times faster than the current best greedy approaches. Reconstruction quality increased up to 45\% with our learned SP over that provided by variable density and Poisson disk SPs, considering the same scan time. Optionally, the scan time can be nearly halved without loss of reconstruction quality. \\\textbf{Conclusion:} Compared with current approaches, BASS can be used to rapidly learn effective SPs for various reconstruction methods, using larger SP and larger datasets. This enables a better selection of efficacious sampling-reconstruction pairs for specific MRI problems.}


\maketitle

\section{Introduction}

\subsection{Background and purpose:}

In magnetic resonance imaging (MRI), the information in the sampled
signal is proportional to the acquisition time \cite{Bernstein2004,Liang-2000}.
This makes the acquisition of high-resolution three-dimensional (3D)
volume imaging of the human body time-consuming. Further, short scan
times in MRI are fundamental for capturing dynamic processes and quantitative
imaging, and in reducing health-care costs.

Fast magnetic resonance (MR) pulse sequences for data acquisition
\cite{Bernstein2004,Liang-2000,Tsao2010}, parallel imaging (PI) using
multichannel receive radio frequency arrays \cite{Ying2010,Pruessmann2006,Blaimer2004},
simultaneous multi-slice imaging (SMS)/multi-band excitation \cite{Feinberg2013,Barth2016},
and compressed sensing (CS) \cite{Lustig-2007,Trzasko-2009,Lustig-2008}
are examples of advancements towards rapid MRI. PI uses multiple receivers
with different spatial coil sensitivities to capture samples in parallel
\cite{Wang2000}, increasing the amount of data in the same scan time.
Consequently, undersampling can be used to reduce the overall scan
time \cite{Ying2010,Pruessmann2006,Blaimer2004}. CS relies on incoherent
sampling and sparse reconstruction. With incoherence, the sparse signals
spread almost uniformly in the sampling domain, and random-like patterns
can be used to undersample the k-space \cite{Lustig-2007,Trzasko-2009,Lustig-2008,Wright2014,Feng2017}. 

Successful reconstructions with undersampled data, such as PI and
CS, use prior knowledge about the true signal to remove artifacts
of undersampling, preserving most of the desired signal. Essentially,
the true signal is redundant and can be compactly represented in a
certain domain, subspace, or manifold, of much smaller dimensionality
\cite{Lee2007,Elad2010a}. Low-rank signal representation \cite{Jacob2020}
and sparse representation \cite{Rubinstein2010}, are two examples
of this kind. Deep learning-based reconstructions have shown that
undersampling artifacts can also be separated from true signals by
learning the parameters of a neural network from sampled datasets
\cite{Jacob2020,Knoll2019a,Liang2020}.

The quality of image reconstruction depends on the sampling process.
CS is an example of how the sampling pattern (SP) should be modified
\cite{Candes2007a,Donoho-2006,Haldar2011}, compared to standard uniform
sampling \cite{Unser-2000}, to be effective for a specific signal
recovery strategy \cite{Haldar2011,Candes-2006b}. According to theoretical
results \cite{Candes2007a,Candes2006a,Donoho2006a}, restricted isometry
properties (RIP) and incoherence are key for CS. In MRI, however,
RIP and incoherence are more like guidelines for designing random
sampling \cite{Lustig-2007,Lustig-2008,Haldar2011} than target properties.
A reason is the difficulty in evaluating such properties for CS methods
in MRI, especially when priors \cite{Elad2007} such as total variation
(TV) \cite{Rudin1992} are utilized. Studies \cite{Zijlstra2016,Boyer2016}
show that SPs with optimally incoherent measurements \cite{Lustig-2007}
do not achieve the best reconstruction quality, leaving room for effective
empirical designs. SPs such as variable density \cite{Cheng2013,Ahmad2015,Wang2010}
or Poisson disk \cite{Dunbar2006,Murphy2012,Kaldate2016} show good
results in MRI reconstruction without having optimal incoherence properties. 

In many CS-MRI methods, image quality improves when SP is learned
utilizing a fully sampled k-space of similar images of particular
anatomy as a reference \cite{Knoll2011a,Choi2016,Vellagoundar2015,Krishna2015,Zhang2014a}.
Such adaptive sampling approaches adjust the probability of the k-space
points of variable density SP according to the k-space energy of reference
images \cite{Knoll2011a,Choi2016,Vellagoundar2015,Krishna2015,Zhang2014a,Kim2015a}.
Such SP design methods have been developed for CS reconstructions,
but generally they do not consider the reconstruction method to be
used.

Statistical methods using optimized design techniques can be used
for finding best sampling patterns\cite{Haldar2019,Seeger2010a}.
Experimental design methods, especially using optimization of Cramer-Rao
bounds, are general and focus on obtaining improved signal-to-noise
ratio (SNR). These approaches were used for fingerprinting\cite{Zhao2019},
PI \cite{Bouhrara2018}, and sparse reconstructions\cite{Haldar2019}.
They do not consider specific capabilities of the reconstruction algorithm
in the design of the SP, even though some general formulation is usually
assumed.

In \textit{data-driven optimization} (DDO) approaches, the SP is optimized
for reference images or datasets containing several images of particular
anatomy, using a specific method for image reconstruction \cite{Gozcu2018,Gozcu2019,Sanchez2020,Duan-duanLiu2012,Ravishankar2011}.
The main premise is that the optimized SP should perform well with
other images of the same anatomy when the same reconstruction method
is used. These approaches can be extended to jointly learning the
reconstruction and the sampling pattern, as shown in \cite{Bahadir2020a,Aggarwal2020a,Weiss2020}.
DDO is applicable to any reconstruction method that accepts various
SPs. In \cite{Gozcu2019}, DDO for PI and CS-MRI is proposed using
greedy optimization of an image domain criterion (an extension of
\cite{Gozcu2018} for single-coil MRI); see also \cite{Sanchez2020}.

Finding an optimal SP in DDO approaches is an NP-hard problem (it
is a subset selection problem \cite{Broughton2010,Zhou2019}). Also,
each candidate SP needs to be evaluated on a large set of k-space
data or images, which may involve reconstructions with high computational
cost. Effective minimization algorithms are fundamental for the applicability
of DDO approaches with large sampling patterns.

\subsection{Existent SP optimization:}

Commonly used in prior works are the greedy approaches; classified
as forward \cite{Gozcu2018,Couvreur1998,Haldar2011} (increase the
number of points sampled in the SP, starting from the empty set),
backward \cite{Haldar2019,Couvreur1998} (reduce the number of points
in the SP, from fully sampled), or hybrid. \cite{Broughton2010} Considering
the current SP, greedy approaches test candidates SPs, that are one
k-space element different, to be added (or removed). After testing,
they add (or remove) the k-space element that provides the best improvement
in the cost function \cite{Zhou2019}.

Greedy approaches have a disadvantage regarding computational cost
because of the large number of evaluations/reconstructions. Assuming
a fully-sampled k-space data is of size $N$, where the undersampled
data is of size $M<N$, and there are $N_{i}$ images, or data items,
used for the learning process, the greedy approach will take $N\times N_{i}$
reconstructions just to find the best first sampling point in the
SP (not considering the next $M-1$ k-space points that still have
to be computed). This makes greedy approaches computationally unfeasible
for large-scale MRI problems. As opposed to this, the approach proposed
in this work can obtain a good SP using $50N_{i}$$\mathrm{\:to\:}$$500N_{i}$
image reconstructions (for all the $M$ k-space points of the SP).

The approach in \cite{Gozcu2018} is only feasible because it was
applied to one-dimensional (1D) undersampling, with a small number
of images in the dataset and single-coil reconstructions. The approach
was extended to 1D+time dynamic sequences\cite{Sanchez2020} and to
parallel imaging\cite{Gozcu2019}, but it requires too many evaluations,
practically prohibitive for large datasets and large sampling patterns.

A different class of learning algorithms for subset selection \cite{Zhou2019},
not exploited yet by SP learning, use bit-wise mutation, such as Pareto
optimization algorithm for subset selection (POSS) \cite{Zhou2019,Qian2015,Qian2017}.
These learning approaches are less costly per iteration since they
evaluate only one candidate SP and accept it if the cost function
is improved. POSS is not designed for fast convergence, but for achieving
good final results. However, these machine learning approaches can
be accelerated if the changes are done smartly and effectively instead
of randomly.

\subsection{The specific content of this paper:}

We propose a new DDO approach to learn the SP in parallel MRI applications.
Our focus is on Cartesian 3D high-resolution and quantitative MRI.
The proposed method can be applied to any parallel MRI method that
has some freedom in selecting the sampling pattern, like CS and low-rank
approaches. Methods that directly recover the k-space elements, such
as simultaneous auto-calibrating and k-space estimation (SAKE) \cite{Shin2014},
low-rank modeling of local k-space neighborhoods (LORAKS) \cite{Haldar2014},
generic iterative re-weighted annihilation filter (GIRAF) \cite{Ongie2017},
and annihilating filter-based low-rank Hankel matrix approach (ALOHA)
\cite{Jin2016a}, among others, can be used. We tested the proposed
optimization approach for P-LORAKS \cite{Haldar2016} and three different
multi-coil CS approaches with different priors \cite{Zibetti2018}.
The main contribution of the proposed approach is a new learning algorithm,
named bias-accelerated subset selection (BASS), that can optimize
larger sampling patterns, using large datasets, spending significantly
less processing times. A very preliminary description of this work
was previously presented in \cite{Zibetti2020a}.

\section{Theory}

\subsection{\label{sub:Mathematical-specification-of}Specification of our aim:}

It is assumed that there is a set $\Gamma$ of $N$ sample points
and our instrument (a multi-coil MRI scanner using Cartesian sampling)
can provide a measurement for any sample point. An ordering of the
$N$ points of $\Gamma$ is represented by an $N$-dimensional complex-valued
vector $\mathbf{m}$, whose components are the measurements at the
sample points; this is referred to as the \textit{fully-sampled data}. 

Let $\Omega$ be any subset (of size $M$) of $\Gamma$; it is referred
to as a \textit{sampling pattern} (SP). Undersampling at points of
$\Omega$ gives rise to a
\begin{equation}
\bar{\mathbf{m}}=\mathbf{S}_{\Omega}\mathbf{m},\label{eq:model_S-1}
\end{equation}
where $\mathbf{S}_{\Omega}$ is an $M\times N$ matrix, referred to
as the \textit{sampling function}. The \textit{acceleration factor}
(AF) is defined as $N/M$.

It is assumed that we have a fixed \textit{recovery algorithm $R$}
that, for any SP $\Omega$ and any undersampled measurements $\bar{\mathbf{m}}$
for that SP, provides an estimate, denoted by $R(\Omega,\bar{\mathbf{m}})$,
of all the measurements. The recovery algorithm is fixed and the $\bar{\mathbf{m}}$
is provided by the instrument, but choice exists for the SP. A method
for finding an efficacious choice in a particular application area
is our subject matter. Efficacy may be measured in the following data-driven
manner.

Let $N_{i}$ be the number of fully sampled data items (denoted by
$\mathbf{m}_{1},\ldots,\mathbf{m}_{N_{i}}$, also called training
data) used in the learning process to obtain an efficacious $\Omega$.
Intuitively, we wish to find a SP $\Omega$ such that all the measurements
$\mathbf{m}_{i}$, for $1\leq i\leq N_{i}$, are ``near'' to their
respective recovered versions $R(\Omega,\mathbf{S}_{\Omega}\mathbf{m}_{i})$
from the undersampled data. Using $f(\mathbf{m},\mathbf{n})$ to denote
the ``distance'' between two fully-sampled measurement vectors $\mathbf{m}$
and $\mathbf{n}$, we define the \textit{efficacy} of an SP $\Omega$
as:
\begin{equation}
F(\Omega)=\frac{1}{N_{i}}\sum_{i=1}^{N_{i}}f\left(\mathbf{m}_{i},R\left(\Omega,\mathbf{S}_{\Omega}\mathbf{m}_{i}\right)\right).\label{eq:DDO-general-1}
\end{equation}
 Then the sought-after \textit{optimal sampling pattern} of size $M$
is:
\begin{equation}
\hat{\Omega}=\argmin_{\underset{s.t.\,|\Omega|=M}{\Omega\subset\Gamma}}F(\Omega).\label{eq:DDO-general}
\end{equation}

\subsection{Models used:}

Parallel MRI methods that directly reconstruct the images, such as
sensitivity encoding method (SENSE) \cite{Pruessmann1999,Pruessmann2006}
and many CS approaches \cite{Zibetti2018c}, are based on an image-to-k-space
forward model, such as
\begin{equation}
\mathbf{m}=\mathbf{\mathbf{FC}x}=\mathbf{Ex},\label{eq:model_E}
\end{equation}
where $\mathbf{x}$ represents a 2D+time image of size $N_{x}\text{\texttimes}N_{y}\text{\texttimes}N_{t}$
($N_{x}$ and $N_{y}$ are horizontal and vertical dimensions, $N_{t}$
is the number of time frames), $\mathbf{C}$ denotes the coil sensitivities
transform mapping $\mathbf{x}$ into multi-coil-weighted images of
size $N_{x}\text{\texttimes}N_{y}\text{\texttimes}N_{t}\text{\texttimes}N_{c}$,
with $N_{c}$ coils.\textcolor{blue}{{} }Each component of $\mathbf{m}$
is an $N_{c}$-dimensional vector. $\mathbf{F}$ represents the spatial
Fourier transforms (FT), comprising $N_{t}\text{\texttimes}N_{c}$
repetitions of the 2D-FT, and $\mathbf{m}$ is the fully sampled data,
of size $N_{x}\times N_{y}\text{\ensuremath{\times}}N_{t}\text{\ensuremath{\times}}N_{c}$.
The two transforms combine into the encoding matrix $\mathbf{E}$.
When accelerated MRI by undersampling is used, the sampling pattern
is included in the model as
\begin{equation}
\bar{\mathbf{m}}=\mathbf{S}_{\Omega}\mathbf{\mathbf{FC}x},\label{eq:model_SE}
\end{equation}
where $\mathbf{S}_{\Omega}$ is the sampling function using SP $\Omega$
(same for all coils) and $\bar{\mathbf{m}}$ is the undersampled multi-coil
k-space data (or k-t-space when $N_{t}>1$), with $M\text{\ensuremath{\times}}N_{c}$
elements. Recall that $M$ is the number of sampled points in the
undersampled k-space and the AF is $N/M$. For reconstructions based
on this model, we assumed that a central area of the k-space is fully
sampled (such an area is used to compute coil sensitivities with auto-calibration
methods, as in \cite{Uecker2014}).

In parallel MRI methods that recover the multi-coil k-space directly,
the undersampling formulation is given by (\ref{eq:model_S-1}) and
the image-to-k-space forward model is not used, since one is interested
in recovering missing k-space samples using e.g. structured low-rank
models \cite{Jacob2020}. For this, the multi-coil k-space is lifted
into a matrix $\mathbf{H}=H(\mathbf{m})$, assumed to be a low-rank
structured matrix. Lifting operators $H(\mathbf{m})$ are slightly
different across PI methods, exploiting different kinds of low-rank
structure \cite{Jacob2020,Shin2014,Haldar2014,Ongie2017,Jin2016a,Haldar2016}.

Once all the samples of the k-space are recovered, the image can be
computed by any coil combination \cite{Walsh2000,Roemer1990}, such
as:
\begin{equation}
[\hat{\mathbf{x}}]_{n}={\textstyle \sum_{c=1}^{N_{c}}}\mathbf{w}_{n,c}[\mathbf{F}_{c}^{-1}\mathbf{m}_{c}]_{n},\label{eq:coil_comb}
\end{equation}
where $\mathbf{m}_{c}$ is the data from coil $c$, $\mathbf{F}_{c}^{-1}$
is the inverse 2D-FT for one coil and $\mathbf{w}_{n,c}$ is the weight
for spatial position $n$ and coil $c$.

\subsection{Reconstruction methods tested:}

We tested our proposed approach on four different reconstruction methods:
Two one-frame parallel MRI methods (P-LORAKS \cite{Haldar2016}, and
PI-CS with anisotropic TV \cite{Liu2008,Liang2009}) and two multi-frame
low-rank and PI-CS methods for quantitative MRI \cite{Zibetti2018}. 

In P-LORAKS \cite{Haldar2016,Haldar2015} the recovery from $\bar{\mathbf{m}}$
produces:
\begin{equation}
R(\Omega,\bar{\mathbf{m}})=\underset{\underset{s,t.\mathbf{S_{\Omega}}\mathbf{m=\bar{\mathbf{m}}}}{\mathbf{m}}}{\argmin}\left\Vert H_{s}(\mathbf{m})-H_{s,r}(\mathbf{m})\right\Vert _{F}^{2},\label{eq:P-LORAKS}
\end{equation}
where the operator $H_{s}(\mathbf{m})$ produces a low-rank matrix
and $H_{s,r}(\mathbf{m})$ produces a hard threshold version of the
same matrix. P-LORAKS exploits consistency between the sampled k-space
data and reconstructed data; it does not require a regularization
parameter. Further, it does not need pre-computed coil sensitivities,
nor fully sampled k-space areas for auto-calibration.

The CS or low-rank (LR) reconstruction \cite{Zibetti2018} is given
by:
\begin{equation}
\hat{\mathbf{x}}=\underset{\mathbf{x}}{\argmin}\left(\left\Vert \bar{\mathbf{m}}-\mathbf{S}_{\Omega}\mathbf{\mathbf{E}x}\right\Vert _{2}^{2}+\lambda P(\mathbf{x})\right)\approx R_{\mathbf{x}}(\Omega,\bar{\mathbf{m}}),\label{eq:CS}
\end{equation}
where $\lambda$ is a regularization parameter. We looked at the regularizations:
$P(\mathbf{x})=\left\Vert \mathbf{Tx}\right\Vert _{1}$, with $\mathbf{T}$
the spatial finite differences (SFD); and low rank (LR), using nuclear-norm
of $\mathbf{x}$ reordered as a Casorati matrix $P(\mathbf{x})=\left\Vert \mathbf{M}(\mathbf{x})\right\Vert _{*}$
\cite{Liang2007}.

CS approaches using redundant dictionaries $\mathbf{D}$ in the synthesis
models \cite{Rubinstein2010,Elad2007}, given by $\mathbf{x}=\mathbf{Du}$,
can be written as:
\begin{equation}
\hat{\mathbf{x}}=\mathbf{\mathbf{D}}\cdot\underset{\mathbf{u}}{\argmin}\left(\left\Vert \bar{\mathbf{m}}-\mathbf{S}_{\Omega}\mathbf{\mathbf{ED}u}\right\Vert _{2}^{2}+\lambda\left\Vert \mathbf{u}\right\Vert _{1}\right)\approx R_{\mathbf{x}}\left(\Omega,\bar{\mathbf{m}}\right).\label{eq:CS-DIC}
\end{equation}

A dictionary to model exponential relaxation processes, like $\text{T}_{2}$
and $\text{T}_{1\rho}$, in MR relaxometry problems is the multi-exponential
dictionary \cite{Zibetti2018,Doneva2010}. It generates a multicomponent
relaxation decomposition\cite{Zibetti2020}. The approximately-equal
symbol $\approx$ is used in (\ref{eq:CS}) and (\ref{eq:CS-DIC}),
since the iterative algorithm for producing $R_{\mathbf{x}}(\Omega,\bar{\mathbf{m}})$,
MFISTA-VA \cite{Zibetti2019a} in this paper, may stop before reaching
the minimum.

\subsection{Criteria utilized in this paper:}

We work primarily with a criterion defined in the multi-coil k-space;
see (\ref{eq:DDO-general-1}) and (\ref{eq:DDO-general}). This criterion
is used by parallel MRI methods that recover the k-space components
directly in a k-space interpolation fashion (and not in the image-space),
such as P-LORAKS \cite{Haldar2016} and others \cite{Jacob2020,Knoll2019a,Shin2014}.
Unless stated otherwise, the $f(\mathbf{m},\mathbf{n})$ in (\ref{eq:DDO-general-1})
is
\begin{equation}
f(\mathbf{m},\mathbf{n})=\dfrac{\left\Vert \mathbf{m}-\mathbf{n}\right\Vert _{2}^{2}}{\left\Vert \mathbf{m}\right\Vert _{2}^{2}}.\label{eq:DDO-proposed}
\end{equation}
The term $\left\Vert \mathbf{m}\right\Vert _{2}^{2}$ normalizes the
error, so that the cost function will not be dominated by datasets
with a strong signal.

For image-based reconstruction methods (e.g., SENSE and multi-coil
CS) using the model in (\ref{eq:model_E}), the $R\left(\Omega,\mathbf{S}_{\Omega}\mathbf{m}_{i}\right)$
in (\ref{eq:DDO-general-1}) is replaced by $\mathbf{E}R_{\mathbf{x}}\left(\Omega,\mathbf{S}_{\Omega}\mathbf{m}_{i}\right)$,
as defined, e.g.\emph{,} in (\ref{eq:CS}) and (\ref{eq:CS-DIC}).
The approach used to obtain the coil sensitivity is part of the method. 

Note that (\ref{eq:DDO-general}) can be modified for image-domain
criteria as well, such as:

\begin{equation}
\hat{\Omega}=\argmin_{\underset{s.t.|\Omega|=M}{\Omega\subset\Gamma}}\left(\frac{1}{N_{i}}\sum_{i=1}^{N_{i}}g\left(\mathbf{x}_{i},R_{\mathbf{x}}(\Omega,\mathbf{S}_{\Omega}\mathbf{m}_{i})\right)\right),\label{eq:DDO-image-domain}
\end{equation}
where $g\left(\mathbf{x},\mathbf{y}\right)$ is a measurement of the
distance between images $\mathbf{x}$ and $\mathbf{y}$. In this case,
the fully-sampled reference must be computed using a reconstruction
algorithm, such as $\mathbf{x}_{i}=R_{\mathbf{x}}(\Gamma,\mathbf{m}_{i})$,
and so it is dependent on to the parameters used in that algorithm.

\subsection{Proposed Data-Driven Optimization:}

Due to the high computational cost of greedy approaches for large
SPs and the relatively low cost of predicting points that are good
next candidates, we propose a new learning approach, similar to POSS
\cite{Zhou2019,Qian2015,Qian2017}, but with new heuristics that significantly
accelerates the subset selection. For a general description of POSS
see \cite[Algorithm 14.2]{Zhou2019}.

Similarly to POSS, the elements to be changed are selected randomly.
Differently from POSS, two heuristic rules, named the measure of importance
(MI) and the positional constraints (PCs), are used to bias in the
selection of the elements with the intent to accelerate convergence.
This is why the algorithm is named bias-accelerated subset selection
(BASS). The MI (defined explicitly in (\ref{eq:error-computation}))
is a weight assigned to each element, indicating how much it is likely
to contribute to decreasing the cost function. The PCs are positional
rules for avoiding selecting in the same iteration elements that may
not provide an additional contribution.

BASS, aims at finding (an approximation of) the $\hat{\Omega}$ of
(\ref{eq:DDO-general}), is described in Algorithm \ref{alg:proposed_algorithm}.
It uses the following user-defined items:
\begin{itemize}
\item $\Omega_{init}$ is the initial SP for the algorithm. It may be any
SP (a Poisson disk, a variable density or even empty SP).
\item $L$ is the number of iterations in the training process.
\item $N$ is the number of points in the fully-sampled data.
\item $M$ is the desired size of the SP ($M<N$).
\item $K_{init}$ is the maximum (initial) number of elements to be added/removed
per iteration ($K_{init}<\min(M,N-M)$).
\item $\rho_{r}$ is a function of two positive-integer variables $K$ and
$M$ ($K<M$), such that $K/M<\rho_{r}(K,M)\leq1.$
\item $\rho_{a}$ is a function of the positive-integer variables $K$,
$M$ and $N$ ($K<N-M$), such that $K/(N-M)<\rho_{a}(K,M,N)\leq1$.
\item \textbf{select-remove$\left(\Omega,K,\rho_{r}(K.M)\right)$ }is a
subset of $\Omega,$ specified below.
\item \textbf{select-add$\left(\Omega,K,\rho_{a}(K,M,N)\right)$ }is a subset
of $\Gamma\backslash\Omega,$ specified below.
\item $F$ is an efficacy function; see (\ref{eq:DDO-general-1}) with the
following.

\begin{itemize}
\item $N_{i}$ is the number of items in the training set.
\item $\mathbf{m}_{1},\ldots,\mathbf{m}_{N_{i}}$ are the data items in
the training set.
\item $R$ is the recovery algorithm from undersampled data.
\end{itemize}
\item $\alpha$ is a reduction factor for the number of elements to be added/removed
per iteration ($0<\alpha<1$).
\end{itemize}
\begin{algorithm}
\caption{BASS }
\label{alg:proposed_algorithm}

\begin{algorithmic}[1]

\State{$\Omega\leftarrow\Omega_{init}$\textbf{ }}

\State{$K\leftarrow K_{init}$\textbf{ }}

\State{$l\leftarrow1$\textbf{ }}

\State{\textbf{while $l<L$ do} }

\State{\textbf{\enskip{}}$\Omega_{r}\leftarrow$\textbf{select-remove$\left(\Omega,K,\rho_{r}(K,M)\right)$
}}\label{select-remove}

\State{\textbf{\enskip{}}$\Omega_{a}\leftarrow$\textbf{select-add$\left(\Omega,K,\rho_{a}(K,M,N)\right)$
}}\label{select-add-} 

\State{\textbf{\enskip{}}$\Omega'\leftarrow\Omega_{a}\cup\left(\Omega\backslash\Omega_{r}\right)$}\label{Omega'}

\State{\textbf{\enskip{}if }$|\Omega'|\neq M$, \textbf{$\Omega\leftarrow\Omega'$}}

\State{\textbf{\enskip{}if }$|\Omega'|=M$}

\State{\textbf{\enskip{}\qquad{}if }$F\left(\Omega'\right)\leq F(\Omega),$\textbf{
$\Omega\leftarrow\Omega'$}}\label{if-F}

\State{\textbf{\enskip{}\qquad{}else }$K\leftarrow\left\lfloor (K-1)\alpha\right\rfloor +1$}

\State{\textbf{\enskip{}}$l\leftarrow l+1$\textbf{ }}

\State{\textbf{return $\Omega$}}

\end{algorithmic}
\end{algorithm}

\subsection{Selection of elements to add to or remove from the SP:}

Elements of $\Omega_{a}$ and $\Omega_{r}$ are selected by the functions
\textbf{select-add} and \textbf{select-remove} in similar ways, described
in the following paragraphs. First, we point out properties of those
selections that ensure the progress of the learning algorithm toward
finding an SP of $M$ elements. The properties in question are that
if $\Omega_{r}$, $\Omega_{a}$, and $\Omega'$ are obtained by Steps
\ref{select-remove}, \ref{select-add-}, and \ref{Omega'}, respectively,
then
\begin{equation}
\left|\Omega_{a}\right|=\min\left(\max(M+K-|\Omega|,0),K\right),\label{eq:K_a}
\end{equation}
\begin{equation}
\left|\Omega_{r}\right|=\min\left(\max(|\Omega|+K-M,0),K\right),\label{eq:K_b}
\end{equation}
\begin{equation}
\left|\Omega'\right|=\left|\Omega\right|+\left|\Omega_{a}\right|-\left|\Omega_{r}\right|.\label{eq:size-of-Omega'}
\end{equation}
From these properties, it follows that if $\left|\Omega\right|<M$,
then $\left|\Omega_{r}\right|<\left|\Omega_{a}\right|=K$ and if $\left|\Omega\right|>M,$
then $\left|\Omega_{a}\right|<\left|\Omega_{r}\right|=K$. Consequently,
\begin{equation}
\left|\left|\Omega'\right|-M\right|<\left|\left|\Omega\right|-M\right|,\label{eq:getting_to_M}
\end{equation}
if $\left|\Omega\right|\neq M$. On the other hand, if $\left|\Omega\right|=M$,
then $\left|\Omega'\right|=\left|\Omega\right|$. Thus, executing
Algorithm \ref{alg:proposed_algorithm} result in $\left|\Omega\right|$
that are converging to $M$.

We now specify details of \textbf{select-add$\left(\Omega,K,\rho_{a}(K,M,N)\right)$
}and \textbf{select-remove$\left(\Omega,K,\rho_{r}(K,M)\right)$}
in Algorithm \ref{alg:proposed_algorithm} as used in this paper.
For $1\leq i\leq N_{i}$, let $\mathbf{e}_{i}=\mathbf{m}_{i}-R\left(\Omega,\mathbf{S}_{\Omega}\mathbf{m}_{i}\right)$.
By Subsection \ref{sub:Mathematical-specification-of}, each of the
$N$ components of $\mathbf{e}_{i}$ is an $N_{c}$-dimensional vector.
For $1\leq k\leq N$ and $1\leq c\leq N_{c}$, $[\mathbf{e}_{i,c}]_{k}$
denotes the $c$th component of the $k$th component of $\mathbf{e}_{i}$.

For \textbf{select-add}, we define a measure of importance (MI), for
$1\leq k\leq N$, as

\begin{equation}
\boldsymbol{\varepsilon}{}_{k}=\frac{1}{N_{i}N_{c}}\sum_{i=1}^{N_{i}}\frac{\sum_{c=1}^{N_{c}}\left|[\mathbf{e}_{i,c}]_{k}\right|^{2}}{\left\Vert \mathbf{m}_{i}\right\Vert _{2}^{2}},\label{eq:error-computation}
\end{equation}
referred to as the $\boldsymbol{\varepsilon}$-map. The purpose of
\textbf{select-add} is to select a sequence of size given by (\ref{eq:K_a}).
Assuming $\left|\Omega_{a}\right|=K$, \textbf{select-add} selects
$K$ possibly best points from in $\Gamma\backslash\Omega$. First,
an approximately $\rho_{a}\times(N-M)$ number of elements are randomly
pre-selected by Bernoulli trials with $\rho_{a}$ probability. To
have more than $K$ pre-selected points, we need $\rho_{a}>K/(N-M)$.
The probability $\rho_{a}$ is set by the user as a function $\rho_{a}(K,M,N)$,
if it is set too small and there are not enough elements, the algorithm
has to increase its value. The $K$ elements with larger $\varepsilon_{k}$
are chosen from the randomly pre-selected group since they are more
likely to be useful for the aim of (\ref{eq:DDO-general}). According
to the PCs, these $K$ elements should be non-adjacent k-space points
and should not be in complex-conjugated positions. Once an element
from the randomly pre-selected group (beginning with elements of larger
MI) is chosen, any other element with smaller MI satisfying the PC
is excluded from the randomly pre-selected group. PCs are used because
those k-space positions may have their error reduced in the next iteration
once the point is included in the SP. The probability $\rho_{a}$
indirectly controls the bias applied to the selected set. Larger probability
implies less randomness and more bias. 

For \textbf{select-remove} a sequence by (\ref{eq:K_b}) of points
that are in \textbf{$\Omega$} is generated in the same way, but using
$\mathbf{r}{}_{k}$ as a MI, instead of $\varepsilon_{k}$, 

\begin{equation}
\mathbf{r}{}_{k}=\frac{1}{N_{i}N_{c}}\sum_{i=1}^{N_{i}}\frac{\sum_{c=1}^{N_{c}}\left|[\mathbf{e}_{i,c}]_{k}\right|^{2}+\delta}{\sum_{c=1}^{N_{c}}\left|[\mathbf{m}_{i,c}]_{k}\right|^{2}+\delta},\label{eq:weigthed-error-computation}
\end{equation}
for $1\leq k\leq N$, denominated $\mathbf{r}$-map, with $\delta$
a small constant to avoid zero/infinity in the defining of $\mathbf{r}{}_{k}$.
The idea of this MI is that a large reconstruction error in a sampled
k-space point $k$, defined as $\sum_{c=1}^{N_{c}}\left|[\mathbf{e}_{i,c}]_{k}\right|^{2}$,
where the expected quadratic value of the element is relatively small,
defined as $\sum_{c=1}^{N_{c}}\left|[\mathbf{m}_{i,c}]_{k}\right|^{2}$,
renders that point as less important for the SP. The elements of this
sequence comprise $\Omega_{r}$, to be removed from \textbf{$\Omega$}
in the process of composing $\Omega'$. 

The probability of pre-selecting elements for removal should be $\rho_{r}>K/M.$
Our recommended choices for lines \ref{select-add-} and \ref{select-remove}
of Algorithm \ref{alg:proposed_algorithm} are $\rho_{a}(K,M,N)=\rho_{r}(K,M)=K/M$,
resulting in more bias in \textbf{select-add} and more randomness
in \textbf{select-remove.} They were used in our experiments. The
same PCs were used in \textbf{select-add} and \textbf{select-remove}.

\begin{figure}
\centering\includegraphics[width=1\textwidth]{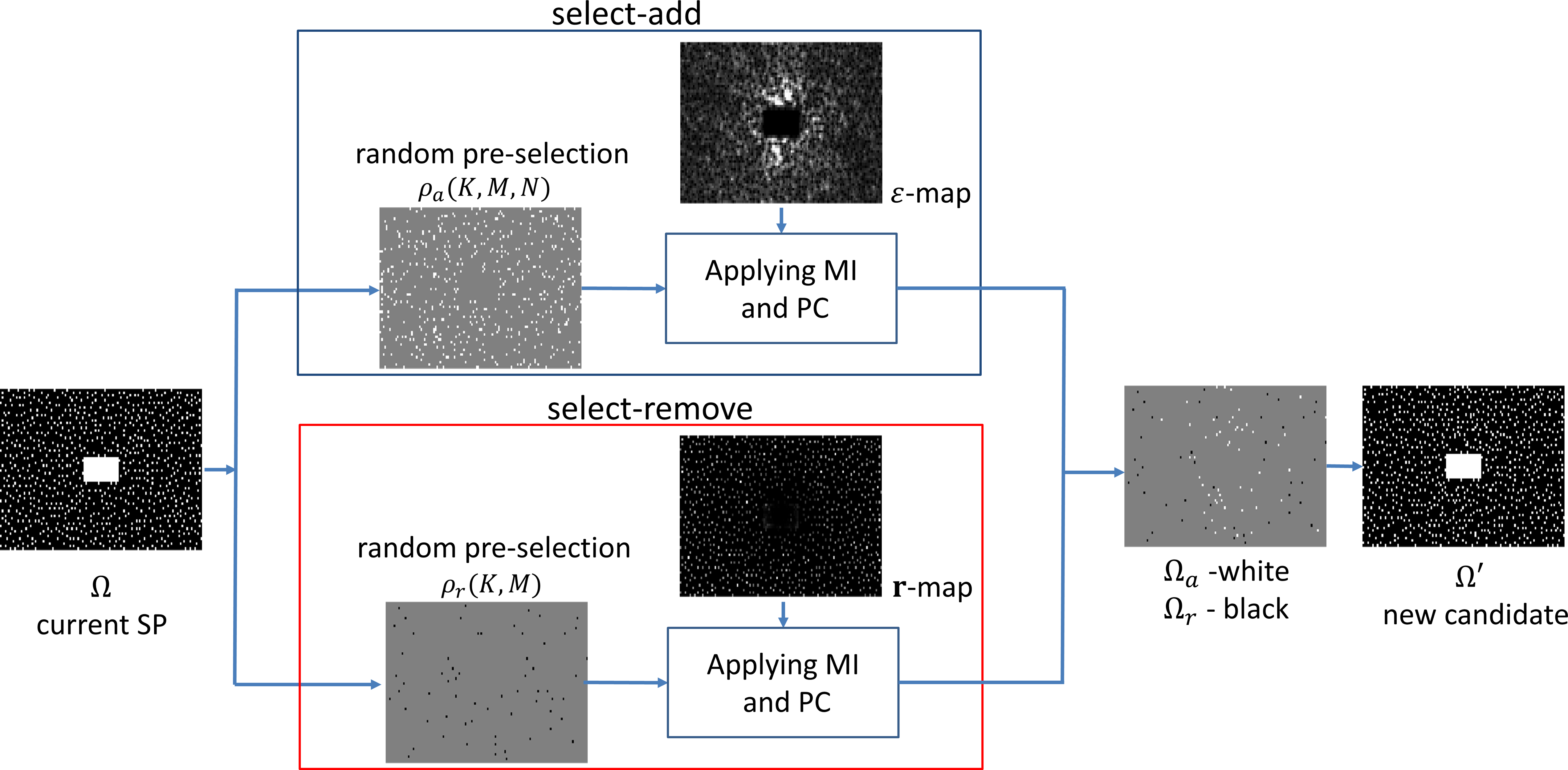}

\caption{Illustration of the steps used in the functions\textbf{ select-add}
and \textbf{select-remove}, first the random pre-selection is done
by Bernoulli trials, using probabilities $\rho_{a}(K,M,N)$ and $\rho_{r}(K,M)$,
and later it is applied the measurement of importance, using $\varepsilon$-map
and $\mathbf{r}$-map (where brighter means higher values), and the
positional constraints. The resulting $\Omega_{a}$ is shown in white
and $\Omega_{r}$ in black in the process of composing $\Omega'$.
The new candidate $\Omega'$ is accepted if the cost function is reduced.
These steps are repeated at each iteration.}
\label{fig:fig-1}
\end{figure}

The expensive part of \textbf{select-add} and \textbf{select-remove}
is the computation of the recoveries given by $R\left(\Omega,\mathbf{S}_{\Omega}\mathbf{m}_{i}\right)$,
but this is done only once per iteration, for all $N_{i}$ images.
Figure \ref{fig:fig-1} illustrates the steps of these functions using
$K=50$.

\section{Methods}

\subsection{Datasets:}

In these experiments, we utilized two datasets. One, denominated \textit{brain},
contains 40 brain $\text{T}_{2}$-weighted images from the fast MRI
dataset of \cite{Knoll2020}. Of these, $N_{i}=30$ were used for
training and $N_{v}=10$ for validation. The k-space data have a size
$N_{x}\text{\texttimes}N_{y}\text{\texttimes}N_{t}\text{\texttimes}N_{c}=320\text{\texttimes}320\text{\texttimes}1\text{\texttimes}16$,
and the reconstructed images are $N_{x}\text{\texttimes}N_{y}\text{\texttimes}N_{t}=320\text{\texttimes}320\text{\texttimes}1$.
The second dataset, denominated \textit{knee}, contains $\text{T}_{1\rho}$-weighted
knee images for quantitative $\text{T}_{1\rho}$ mapping, of size
$N_{x}\text{\texttimes}N_{y}\text{\texttimes}N_{t}\text{\texttimes}N_{c}=128\text{\texttimes}64\text{\texttimes}10\text{\texttimes}15$,
and the reconstructed images are $N_{x}\text{\texttimes}N_{y}\text{\texttimes}N_{t}=128\text{\texttimes}64\text{\texttimes}10$.
Unless otherwise stated, $N_{i}=30$ were used for training and $N_{v}=10$
for validation. The k-space data for all images are normalized by
the largest component. A \textit{reduced-size knee} dataset uses only
part of the knee dataset. Images are of size $N_{x}\text{\texttimes}N_{y}\text{\texttimes}N_{t}=128\text{\texttimes}64\text{\texttimes}1,$
and $N_{i}=5$ and $N_{v}=5$. This dataset is used in experiments
with a large number of iterations to compare BASS with greedy approaches.

\subsection{Reconstruction methods:}

For the brain dataset, two reconstruction methods were used:
\begin{itemize}
\item P-LORAKS \cite{Haldar2016}: from Equation (\ref{eq:P-LORAKS}), with
codes available online (https://mr.usc.edu/download/loraks2/).
\item CS-SFD \cite{Zibetti2019a}: Multi-coil CS with sparsity in the spatial
finite differences (SFD) domain, following equation (\ref{eq:CS}),
and minimized with MFISTA-VA.
\end{itemize}
For the $\text{T}_{1\rho}$-weighted knee dataset, we used different
methods:
\begin{itemize}
\item CS-LR \cite{Zibetti2018}: Multi-coil CS using nuclear-norm of the
vector $\mathbf{x}$ reordered as a Casorati matrix $P(\mathbf{x})=\left\Vert \mathbf{M}(\mathbf{x})\right\Vert _{*}$
and minimized with MFISTA-VA. 
\item CS-DIC \cite{Zibetti2018}: Multi-coil CS using synthesis approach
following equation (\ref{eq:CS-DIC}), using $\mathbf{D}$ as a multi-exponential
dictionary \cite{Doneva2010}, and minimized with MFISTA-VA.
\end{itemize}
CS-SFD, CS-LR, and CS-DIC need a fully-sampled area for auto-calibration
of coil sensitivities using ESPIRiT \cite{Uecker2014}. P-LORAKS does
not use auto-calibration. See \textit{https://cai2r.net/resources/software/data-driven-learning-sampling-pattern}
for the codes.

The regularization parameter (the $\lambda$ in (\ref{eq:CS}) and
(\ref{eq:CS-DIC})) required in CS-SFD, CS-LR, and CS-DIC was optimized
independently for each type of SP (Poisson disk, variable density,
or optimized) and each AF, using the same training data. The parameters
of the recovery method $R$ are assumed to be fixed during the learning
process of the SP and re-optimized for the learned SP. Grid optimization
with 50 realizations of the Poisson disk and variable density was
performed, changing the parameters used to generate these SPs, to
obtain the best realization of these SPs; as in \cite{Gozcu2019}.
Poisson disk and variable density codes used in the experiments are
at \textit{https://github.com/mohakpatel/Poisson-Disc-Sampling} and
\textit{http://mrsrl.stanford.edu/\textasciitilde{}jycheng/software.html}.

\subsection{Evaluation of the error:}

The quality of the results obtained with the SP was evaluated using
the normalized root mean squared error (NRMSE):
\begin{equation}
\text{NRMSE}\left(\left\{ \mathbf{m}_{i}\right\} _{i=1}^{N_{v}},\left\{ \hat{\mathbf{m}}_{i}\right\} _{i=1}^{N_{v}}\right)=\sqrt{{\textstyle \sum_{i=1}^{N_{v}}\frac{\left\Vert \mathbf{m}_{i}-\hat{\mathbf{m}}_{i}\right\Vert _{2}^{2}}{\left\Vert \mathbf{m}_{i}\right\Vert _{2}^{2}}}}.\label{eq:NRMSE}
\end{equation}
When not specified, the NRMSE shown was obtained from k-space on the
validation set; results using image-domain and the training set are
also provided, as is structural similarity (SSIM) \cite{Wang2004a}
in some cases.

\section{Results\label{sec:Results}}

\subsection{Illustration of the convergence and choice of parameters:}

\begin{figure}
\centering

\begin{minipage}[t]{0.45\textwidth}%
\subfloat[]{\includegraphics[width=1\columnwidth]{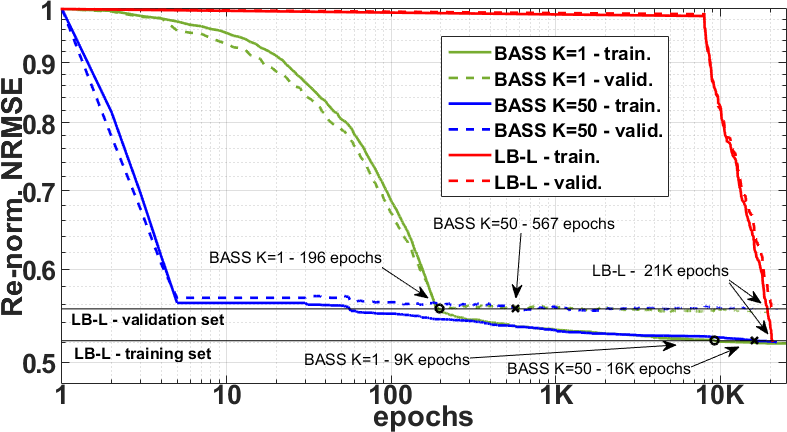}}%
\end{minipage}\,%
\begin{minipage}[t]{0.45\textwidth}%
\subfloat[]{\includegraphics[width=1\columnwidth]{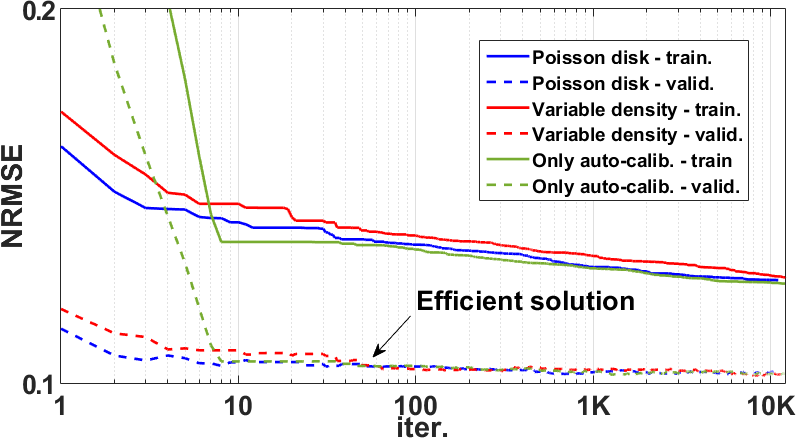}}%
\end{minipage}

\begin{minipage}[t]{0.45\textwidth}%
\subfloat[]{\includegraphics[width=1\columnwidth]{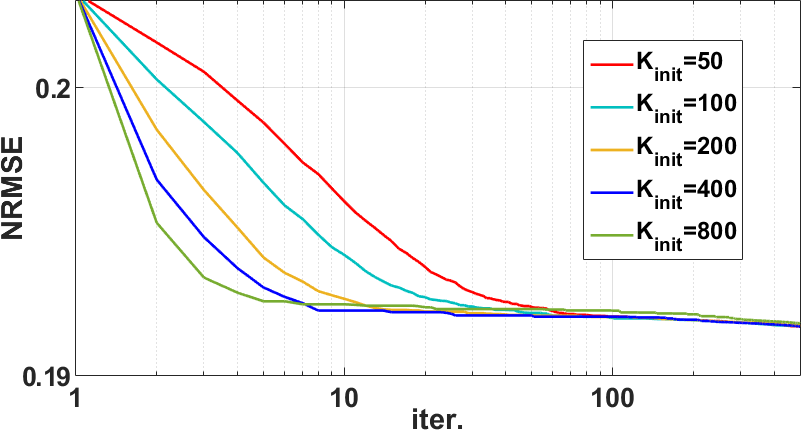}}%
\end{minipage}\,%
\begin{minipage}[t]{0.45\textwidth}%
\subfloat[]{\includegraphics[width=1\columnwidth]{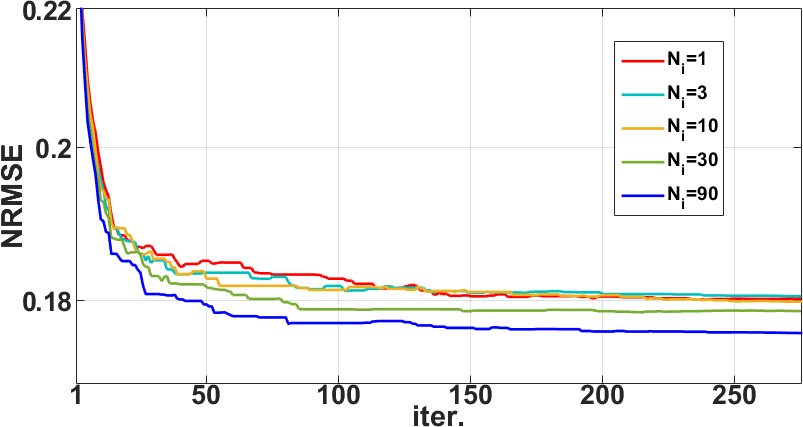}}%
\end{minipage}\caption{Convergence curves for BASS. (a) Comparison with the greedy approach
LB-L. (b) Comparing various initial SPs. (c) Comparing various $K_{init}$s.
(d) Comparing various training sizes.}
\label{fig:fig-2}
\end{figure}

In Figure \ref{fig:fig-2}a we compare BASS with the greedy approach
``learning-based lazy'' (LB-L)\cite{Gozcu2019}, adapted to the
cost function in (\ref{eq:DDO-general-1}). We used CS-SFD with the
\textit{reduced-size knee} dataset, starting with the auto-calibration
area and AF=20. The resulting NRMSEs re-normalized by the initial
values, show the difference in computational cost between the approaches.
Plots are scaled logarithmically in epochs (in each ``epoch'' all
the images are reconstructed once). BASS (K=1, 196 epochs) found an
efficient solution $100$ times faster than LB-L (\textasciitilde{}21,000
epochs). BASS goes on minimizing the cost function beyond the stopping
point of LB-L.

Figure \ref{fig:fig-2}b, demonstrates the performance of BASS for
various initial SPs (same experimental setup as for Figure \ref{fig:fig-2}a,
but using AF=15, and $K_{init}$=50). The improvement observable in
the validation set ends quickly, at iteration 50 in this example.
There is an arrow in the figure pointing to an efficient solution.
Such a solution is obtained after a relatively few iterations, during
which most of the significant improvement observable with validation
data has already happened. Iterating beyond this point essentially
leads to marginal improvement, observable only with the training data.

In Figure \ref{fig:fig-2}c we see the results of the learning process
for the training data according to the parameters $K_{init}$ for
CS-LR, AF=20, using the\textit{ knee} dataset, with $N_{i}=30$ and
$N_{v}=10$. Note that large $K_{init}$ performs better than small
$K_{init}$ in terms of speed of convergence in the beginning of the
learning process. Over time, $K$ reduces from $K_{init}$ towards
$K=1$.

The importance of large and diverse datasets to generate the learned
sampling pattern for the specific class of images is illustrated in
Figure \ref{fig:fig-2}d, showing the convergence of the learning
process with the validation set, in NRMSE. We used training sets of
1, 3, 10, 30, and 90 images. The validation sets were composed of
the same 20 images, not used in any of the training sets.

\begin{figure}
\centering

\begin{minipage}[t]{0.3\columnwidth}%
\subfloat[]{\includegraphics[width=1\columnwidth]{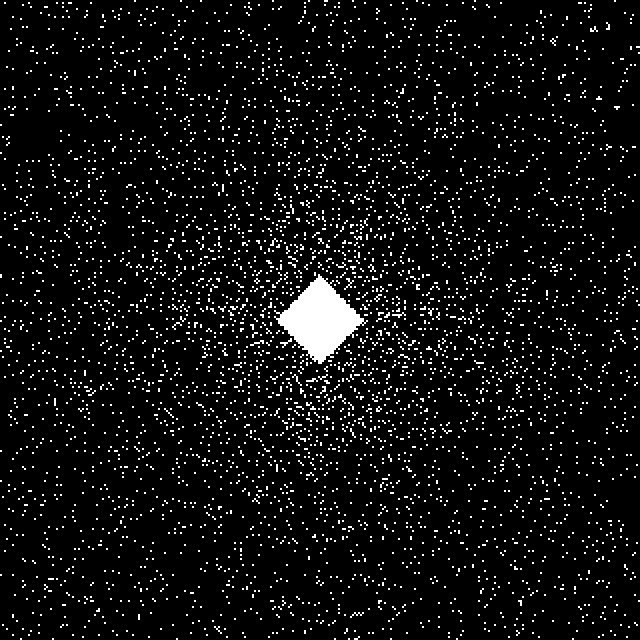}}%
\end{minipage}\,%
\begin{minipage}[t]{0.3\columnwidth}%
\subfloat[]{\includegraphics[width=1\columnwidth]{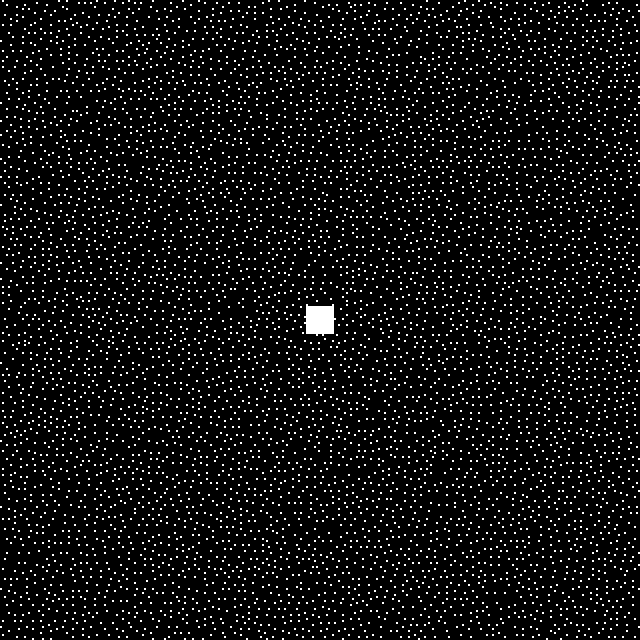}}%
\end{minipage}\,%
\begin{minipage}[t]{0.3\columnwidth}%
\subfloat[]{\includegraphics[width=1\columnwidth]{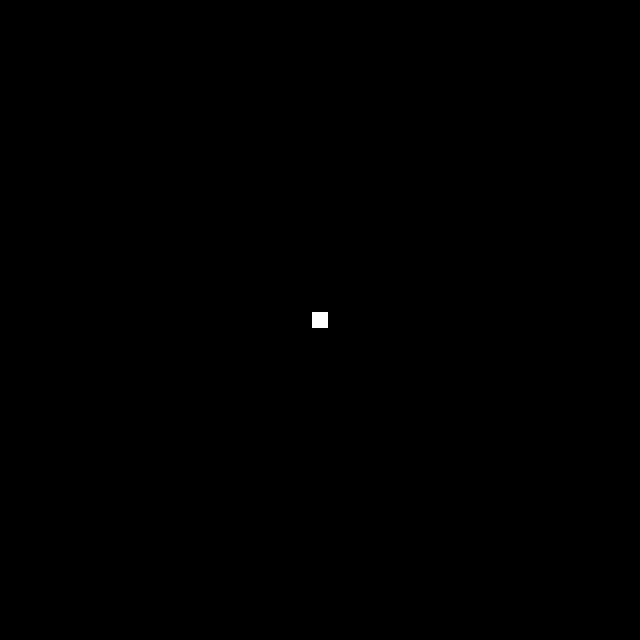}}%
\end{minipage}

\begin{minipage}[t]{0.3\columnwidth}%
\subfloat[]{\includegraphics[width=1\columnwidth]{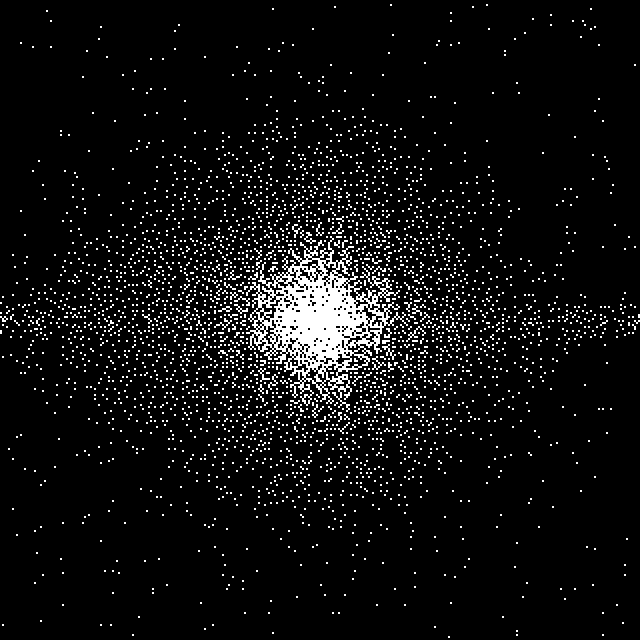}}%
\end{minipage}\,%
\begin{minipage}[t]{0.3\columnwidth}%
\subfloat[]{\includegraphics[width=1\columnwidth]{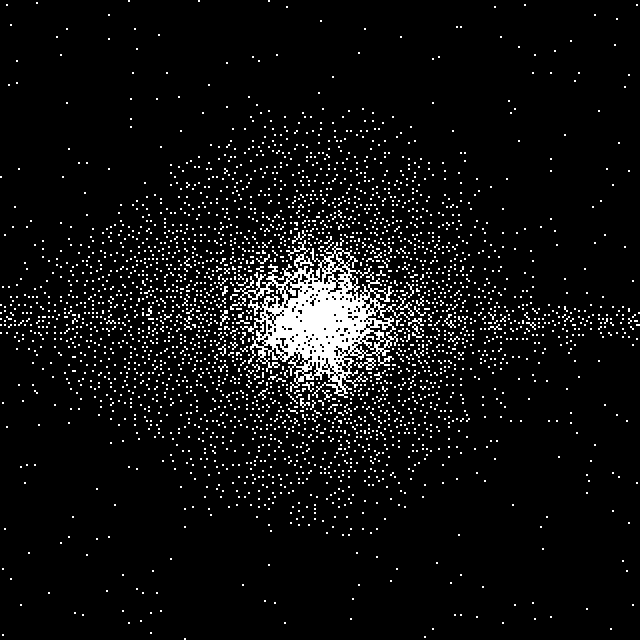}}%
\end{minipage}\,%
\begin{minipage}[t]{0.3\columnwidth}%
\subfloat[]{\includegraphics[width=1\columnwidth]{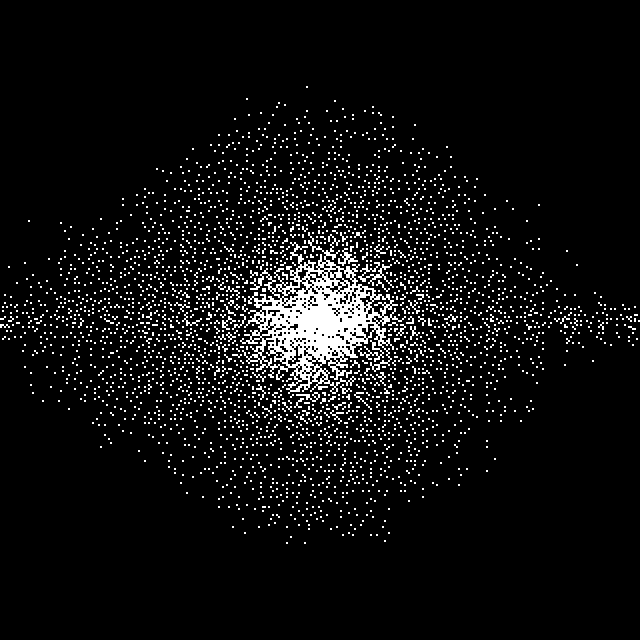}}%
\end{minipage}

\caption{Efficient solutions produced for P-LORAKS with AF=16 and initial SPs
(a) variable density (VD), (b) Poisson disk (PD), and (c) an SP that
is empty except for a small central area (CA). The corresponding efficient
solutions are the SPs in (d) for VD (NRMSE=0.196), in (e) for PD (NRMSE=0.195),
and in (f) for CA (NRMSE=0.194).}
\label{fig:fig-3}
\end{figure}
The robustness of an efficient solution in the presence of variable
initial SP is illustrated in Figure \ref{fig:fig-3}. Figure \ref{fig:fig-3}a-c
show three initial SPs: variable density (VD), Poisson disk (PD),
and empty except for a small central area (CA). Using 200 iterations
of BASS for P-LORAKS with these initial SPs, corresponding efficient
SPs were obtained; shown in Figure \ref{fig:fig-3}d-f. There are
minor differences among them (less than 1\% difference in NRMSE),
but the central parts of the SPs are very similar.

\subsection{Performance with various reconstruction methods:}

\begin{figure}[!th]
\centering

\begin{minipage}[t]{0.45\textwidth}%
\subfloat[]{\includegraphics[width=1\columnwidth]{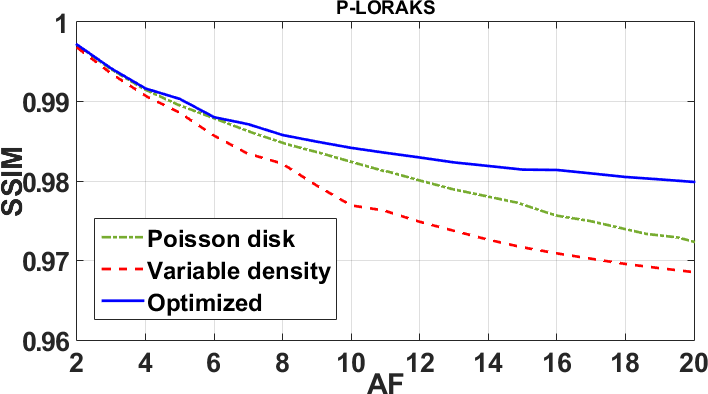}}%
\end{minipage}\,%
\begin{minipage}[t]{0.45\textwidth}%
\subfloat[]{\includegraphics[width=1\columnwidth]{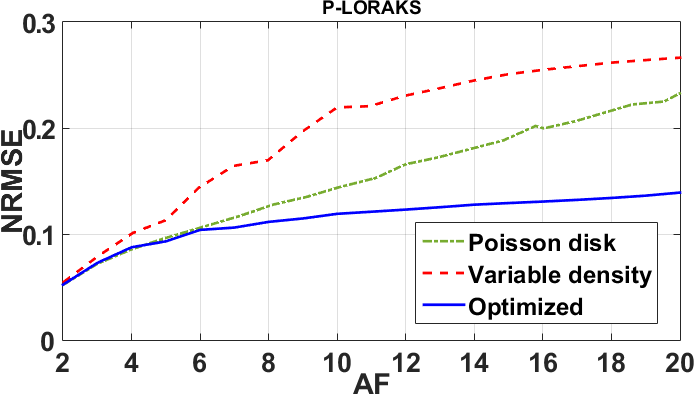}}%
\end{minipage}

\begin{minipage}[t]{0.45\textwidth}%
\subfloat[]{\includegraphics[width=1\columnwidth]{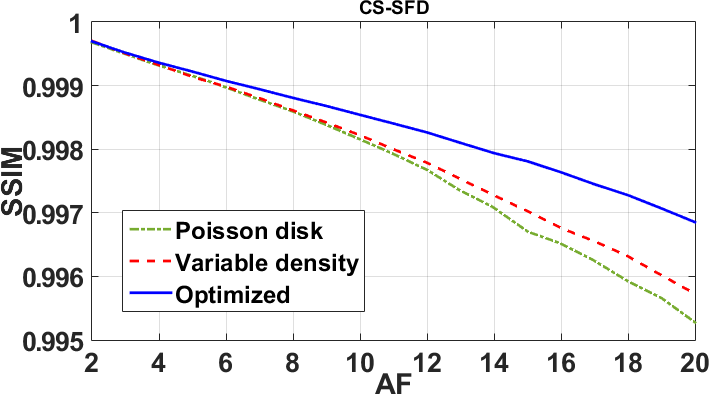}}%
\end{minipage}\,%
\begin{minipage}[t]{0.45\textwidth}%
\subfloat[]{\includegraphics[width=1\columnwidth]{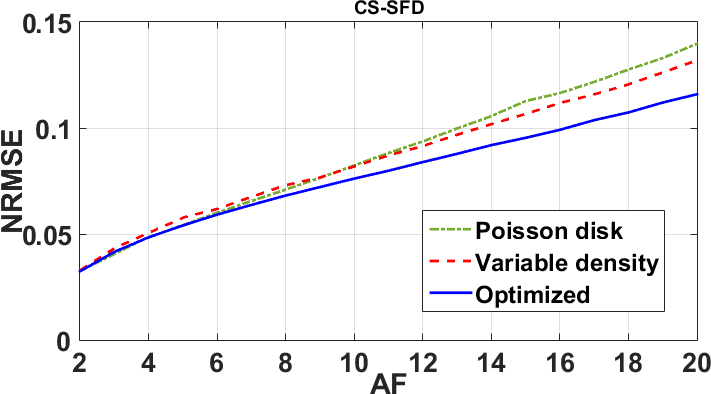}}%
\end{minipage}

\begin{minipage}[t]{0.45\textwidth}%
\subfloat[]{\includegraphics[width=1\columnwidth]{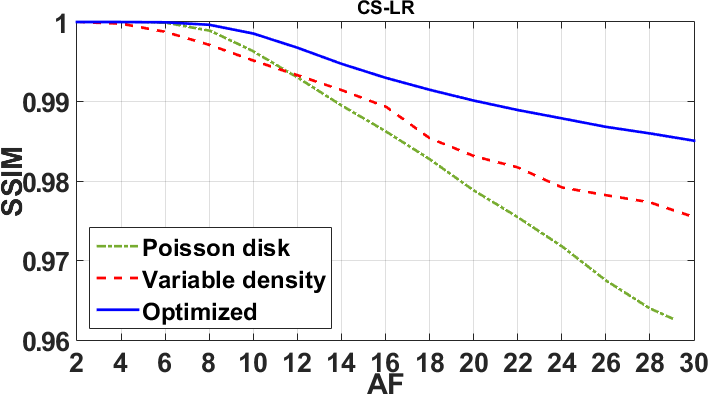}}%
\end{minipage}\,%
\begin{minipage}[t]{0.45\textwidth}%
\subfloat[]{\includegraphics[width=1\columnwidth]{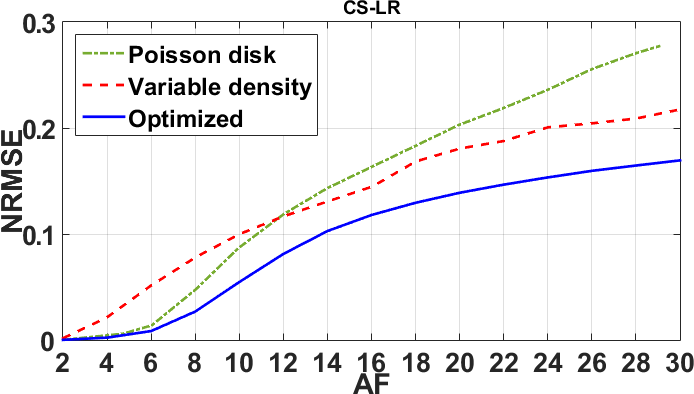}}%
\end{minipage} 

\begin{minipage}[t]{0.45\textwidth}%
\subfloat[]{\includegraphics[width=1\columnwidth]{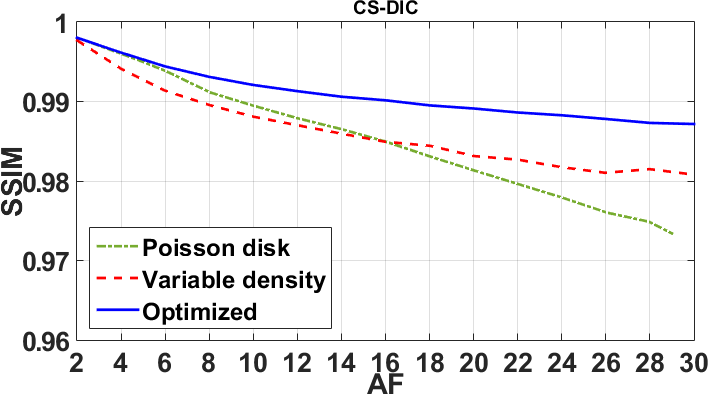}}%
\end{minipage}\,%
\begin{minipage}[t]{0.45\textwidth}%
\subfloat[]{\includegraphics[width=1\columnwidth]{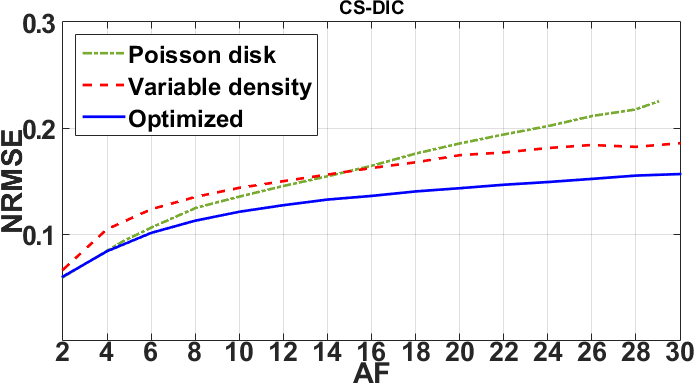}}%
\end{minipage}

\caption{SSIM (higher is better) and NRMSE (lower is better): (a)-(b) for P-LORAKS,
(c)-(d) for CS-SFD, (e)-(f) for CS-LR, and (g)-(h) for CS-DIC, using
variable density or Poisson disk SP compared with optimized SP (obtained
by BASS) for various AFs. (a)-(d) are for \textit{brain} dataset and
(e)-(h) for \textit{knee} dataset.}
\label{fig:fig-5}
\end{figure}

BASS improves both SSIM and NRMSE in image space for fixed AFs when
compared with variable density sampling or Poisson disk for the four
reconstruction methods. Figures \ref{fig:fig-5}a-b show the SSIM
and NRMSE obtained by P-LORAKS with \textit{brain} dataset, compared
to variable density, Poisson disk, and the optimized SP. Figures \ref{fig:fig-5}c-d
show the SSIM and NRMSE obtained by CS-SFD with \textit{brain} dataset.
Figures \ref{fig:fig-5}e-f show the SSIM and NRMSE obtained by CS-LR
with \textit{knee} dataset, comparing Poisson disk, variable density,
and the optimized SP. Figures \ref{fig:fig-5}g-h show the SSIM and
NRMSE obtained by CS-DIC with \textit{knee} dataset. The Poisson disk
and variable density SP had their parameters optimized for each reconstruction
method and each AF.

\begin{figure}
\centering

\begin{minipage}[t]{0.19\textwidth}%
\subfloat[FS]{\includegraphics[width=1\columnwidth]{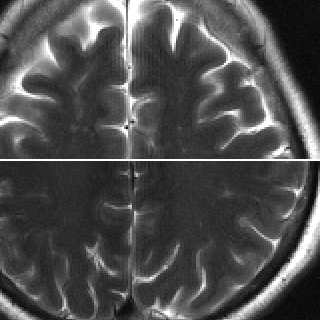}}%
\end{minipage}\,%
\begin{minipage}[t]{0.19\textwidth}%
\subfloat[Optimized AF=16]{\includegraphics[width=1\columnwidth]{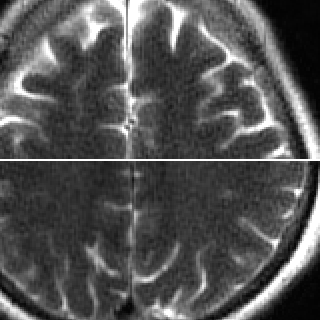}}%
\end{minipage}\,%
\begin{minipage}[t]{0.19\textwidth}%
\subfloat[Poisson disk AF=16]{\includegraphics[width=1\columnwidth]{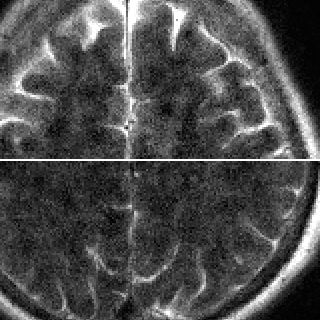}}%
\end{minipage}\,%
\begin{minipage}[t]{0.19\textwidth}%
\subfloat[Optimized AF=16]{\includegraphics[width=1\columnwidth]{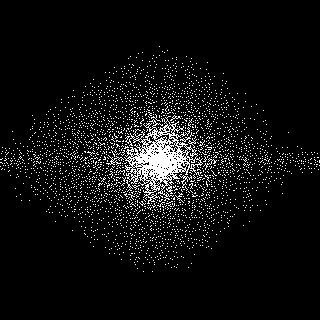}}%
\end{minipage}\,%
\begin{minipage}[t]{0.19\textwidth}%
\subfloat[Poisson disk AF=16]{\includegraphics[width=1\columnwidth]{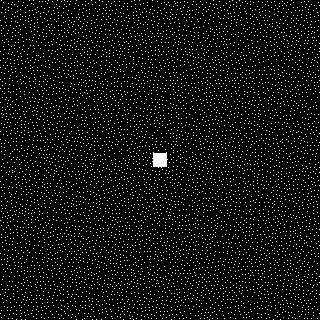}}%
\end{minipage}\quad{}%
\begin{minipage}[t]{0.19\textwidth}%
\subfloat[FS]{\includegraphics[width=1\columnwidth]{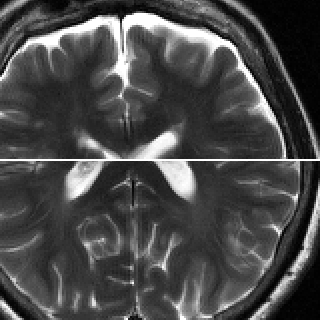}}%
\end{minipage}\,%
\begin{minipage}[t]{0.19\textwidth}%
\subfloat[Optimized AF=16]{\includegraphics[width=1\columnwidth]{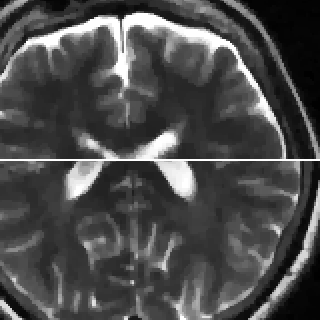}}%
\end{minipage}\,%
\begin{minipage}[t]{0.19\textwidth}%
\subfloat[Variable density AF=16]{\includegraphics[width=1\columnwidth]{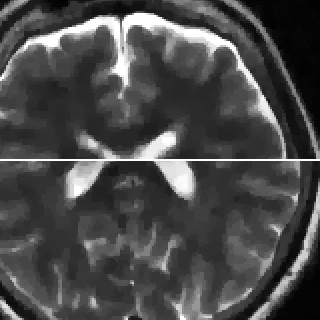}}%
\end{minipage}\,%
\begin{minipage}[t]{0.19\textwidth}%
\subfloat[Optimized AF=16]{\includegraphics[width=1\columnwidth]{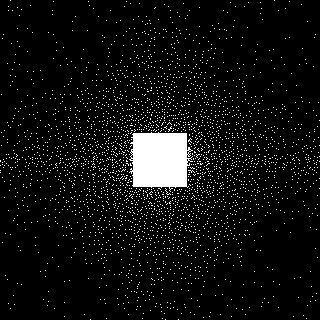}}%
\end{minipage}\,%
\begin{minipage}[t]{0.19\textwidth}%
\subfloat[Variable density AF=16]{\includegraphics[width=1\columnwidth]{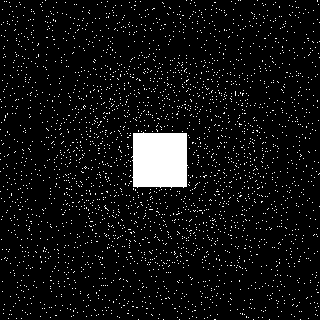}}%
\end{minipage}

\caption{Images of the brain dataset reconstructed with P-LORAKS are shown
in (b)-(c), with CS-SFD in (g)-(h) and using fully sampled (FS) data
in (a) and (f). Optimized and Poisson disk SPs used with P-LORAKS
are shown in (d) and (e) and optimized and variable density SPs used
in CS-SFD, with a central-square auto-calibration region, in (i) and
(j).}
\label{fig:fig-6}
\end{figure}

Figure \ref{fig:fig-6} illustrates on the brain dataset how the optimized
SPs improve the reconstructed images with P-LORAKS and CS-SFD (for
AF=16). The P-LORAKS methods had a visible improvement in SNR, the
CS-SFD methods became less smooth with some structures more detailed.
Figure \ref{fig:fig-6} also illustrates that optimized SPs are different
for the two reconstruction methods, even when using the same images
for training.

\begin{figure}
\centering

\begin{minipage}[t]{0.48\columnwidth}%
\subfloat[CS-DIC Poisson disk]{\includegraphics[width=1\columnwidth]{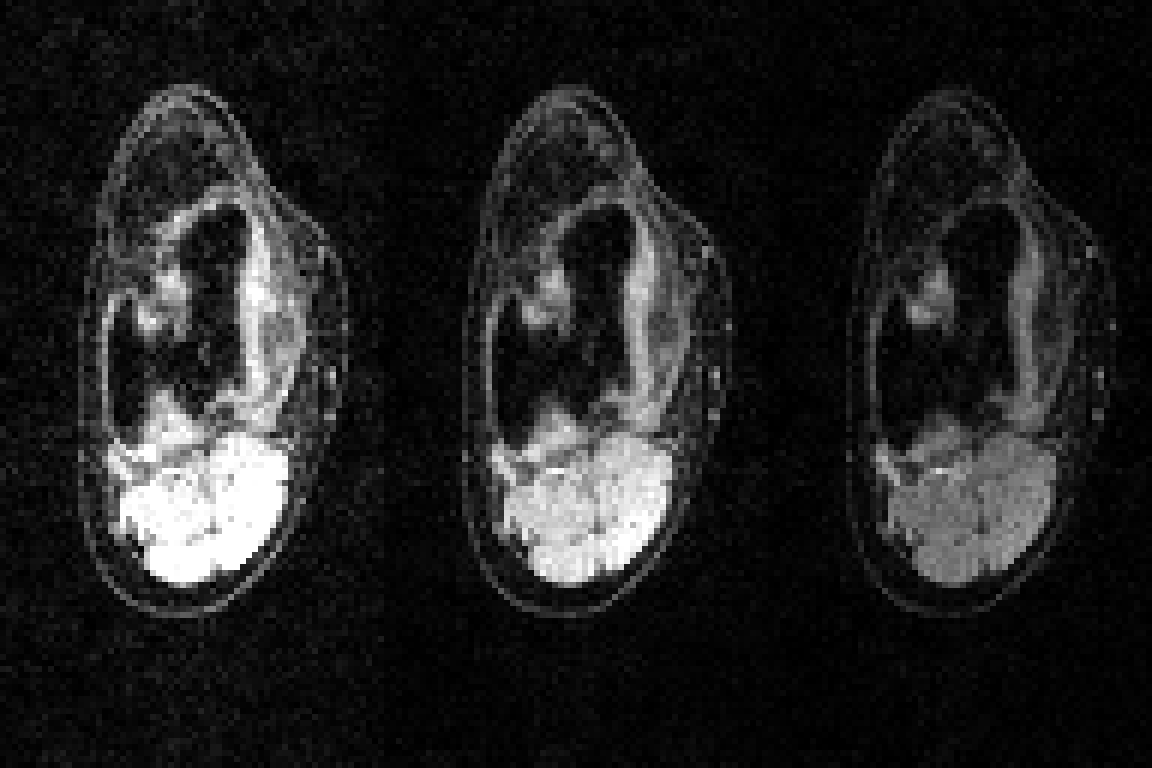}}%
\end{minipage}\,%
\begin{minipage}[t]{0.48\columnwidth}%
\subfloat[CS-DIC Optimized]{\includegraphics[width=1\columnwidth]{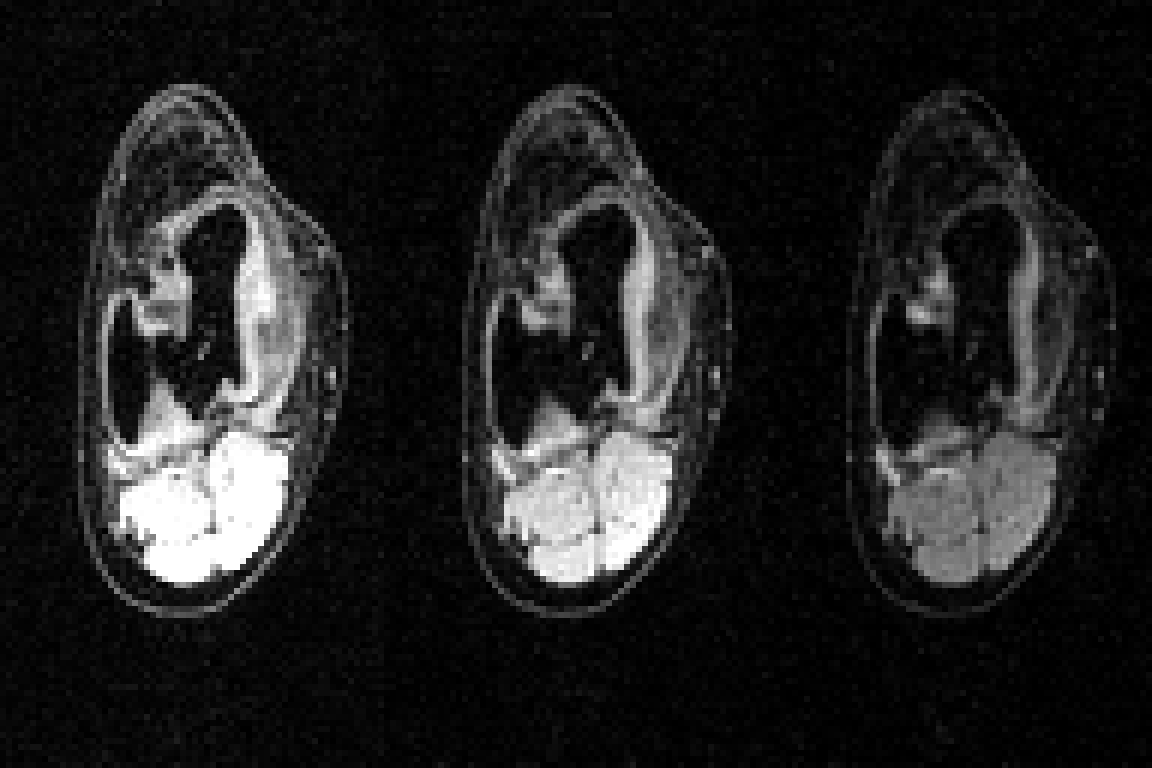}}%
\end{minipage}\enskip{}%
\begin{minipage}[t]{0.48\columnwidth}%
\subfloat[Poisson disk SP]{\includegraphics[width=1\columnwidth]{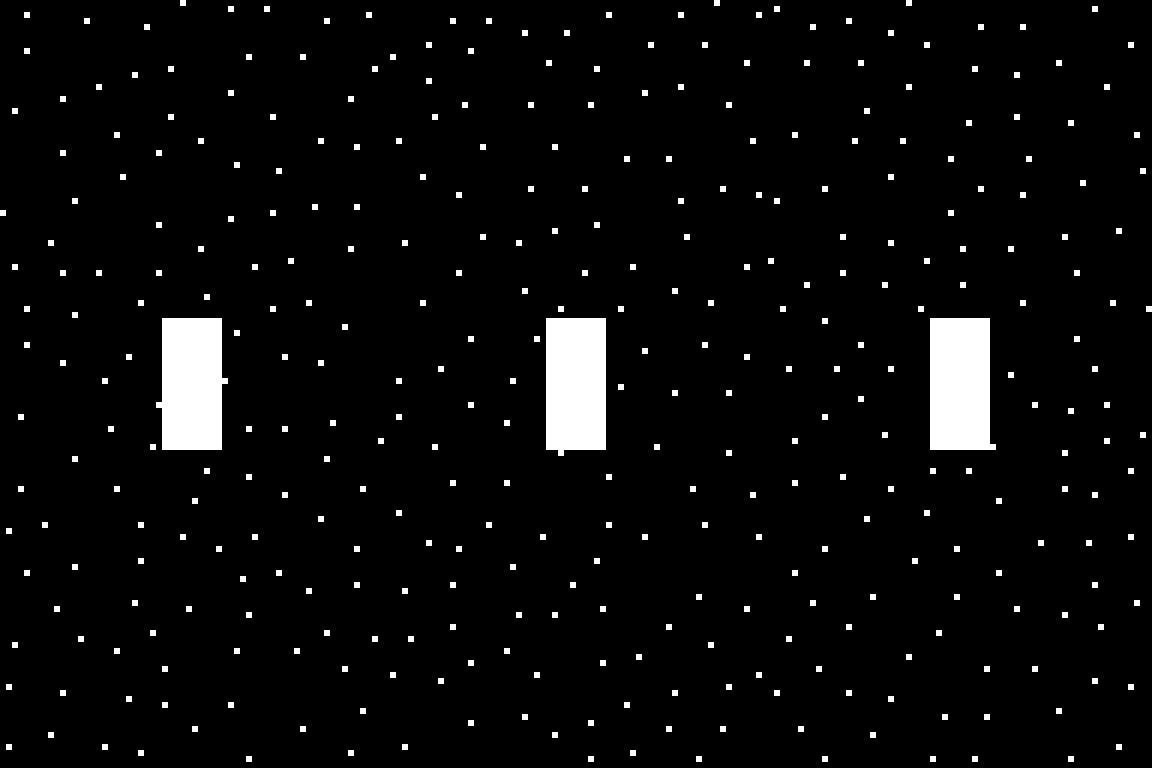}}%
\end{minipage}\,%
\begin{minipage}[t]{0.48\columnwidth}%
\subfloat[Optimized SP CS-DIC]{\includegraphics[width=1\columnwidth]{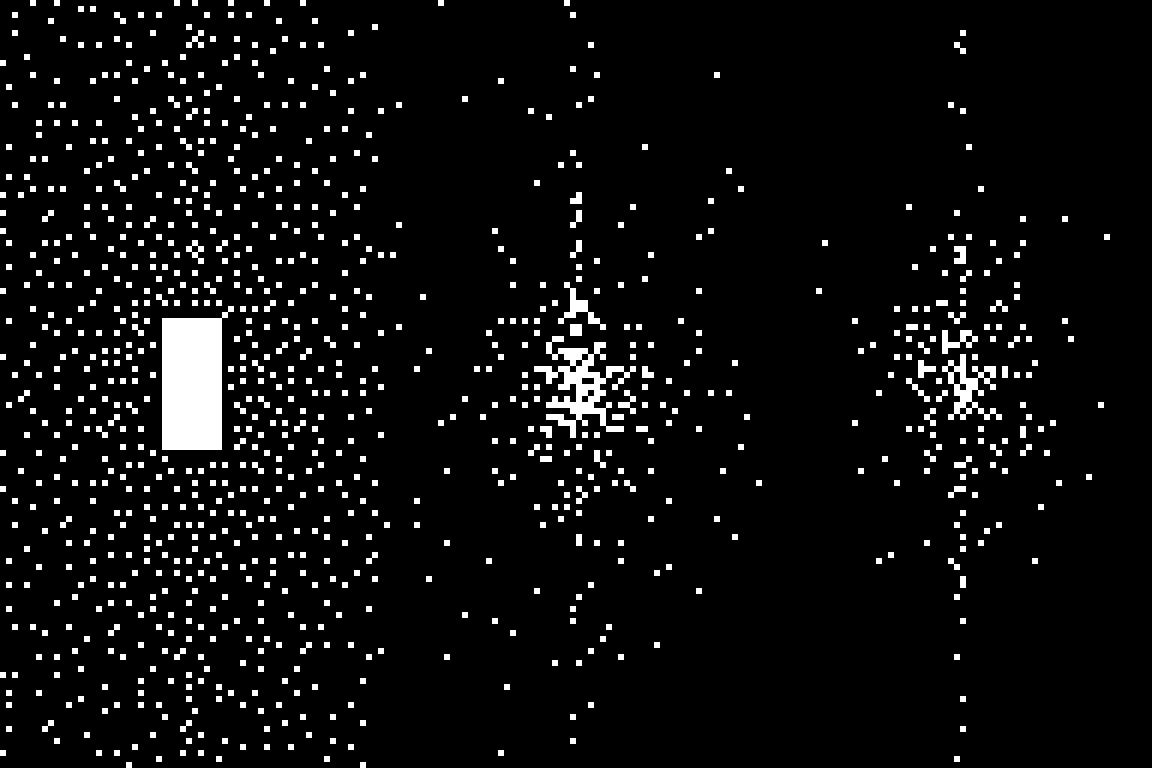}}%
\end{minipage}\enskip{}%
\begin{minipage}[t]{0.48\columnwidth}%
\subfloat[CS-LR Poisson disk]{\includegraphics[width=1\columnwidth]{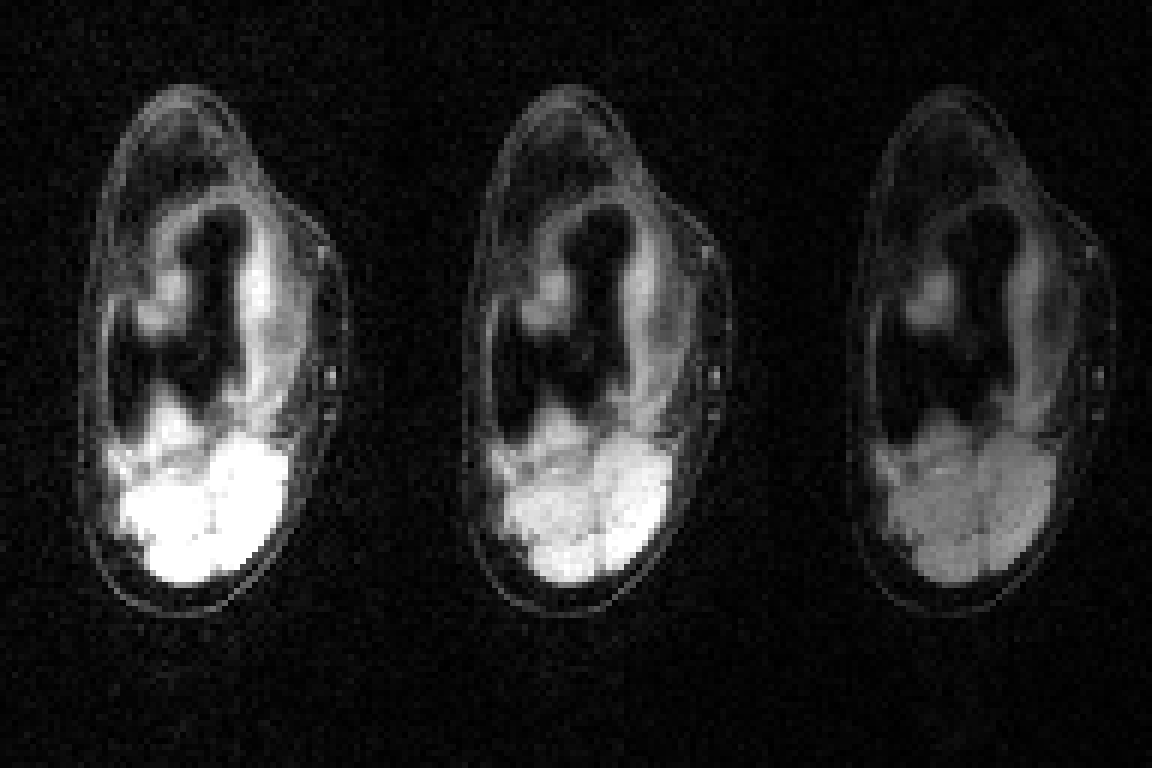}}%
\end{minipage}\,%
\begin{minipage}[t]{0.48\columnwidth}%
\subfloat[CS-LR Optimized]{\includegraphics[width=1\columnwidth]{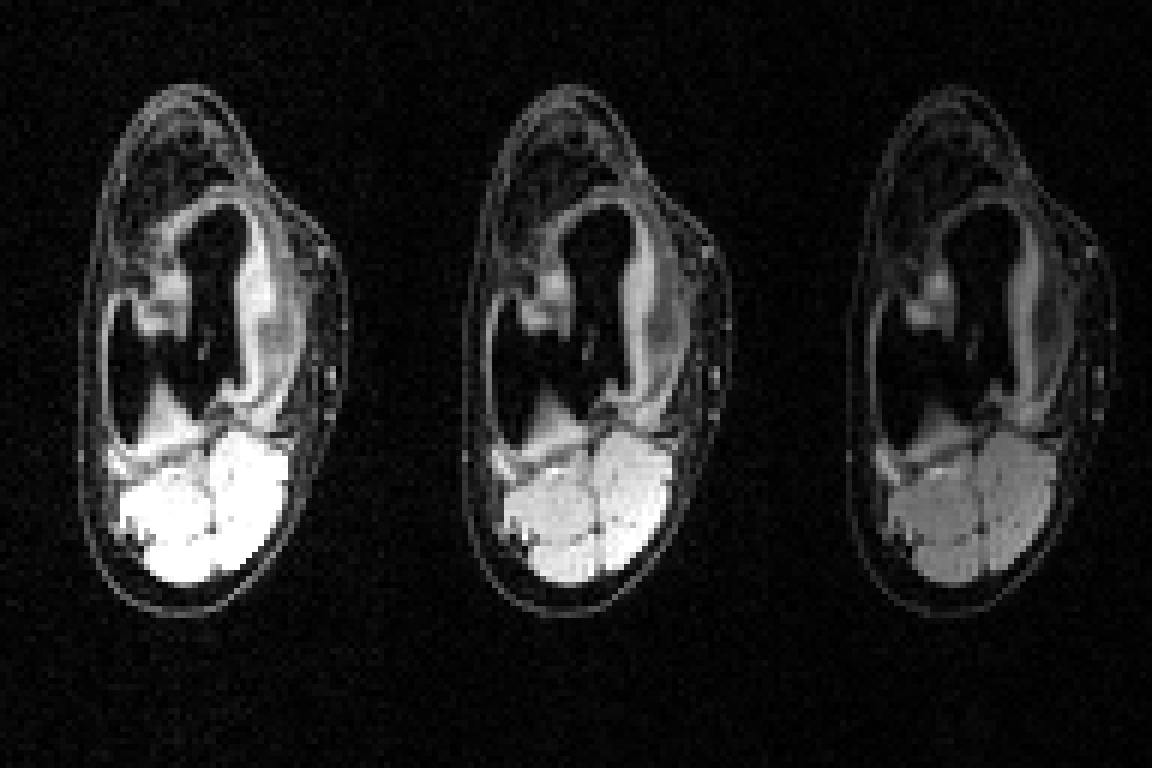}}%
\end{minipage}\enskip{}%
\begin{minipage}[t]{0.48\columnwidth}%
\subfloat[FS]{\includegraphics[width=1\columnwidth]{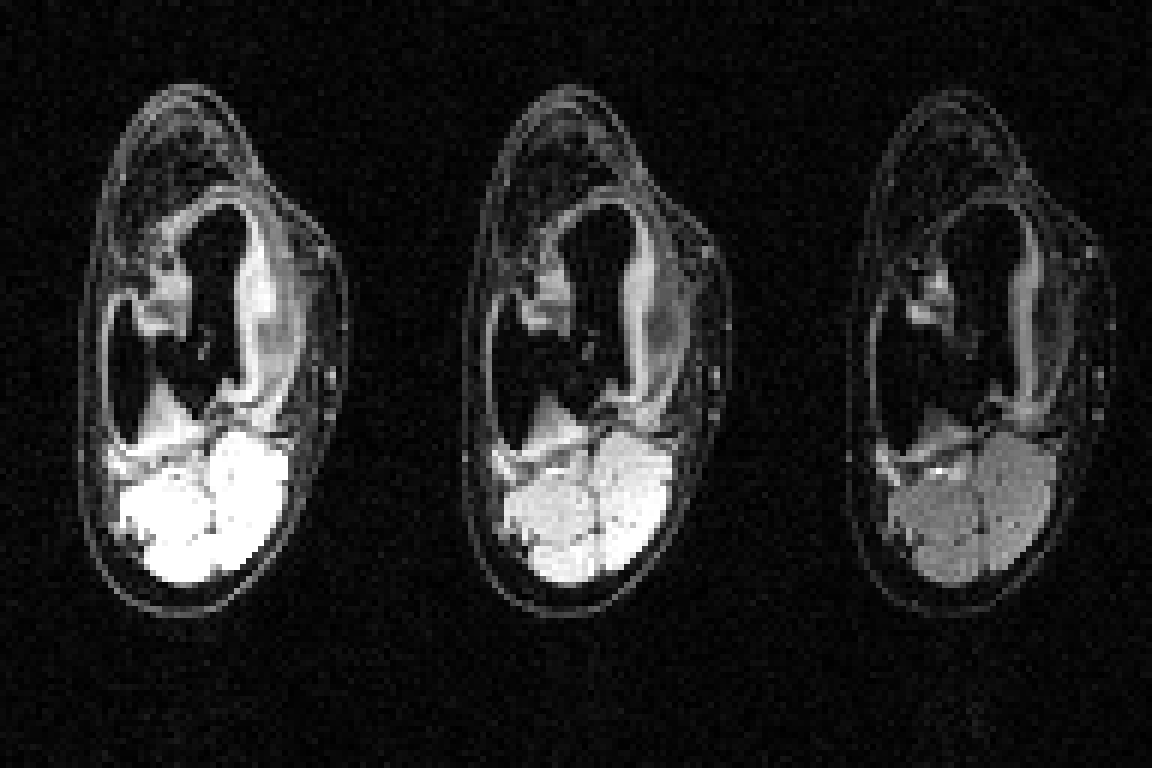}}%
\end{minipage}\,%
\begin{minipage}[t]{0.48\columnwidth}%
\subfloat[Optimized SP CS-LR]{\includegraphics[width=1\columnwidth]{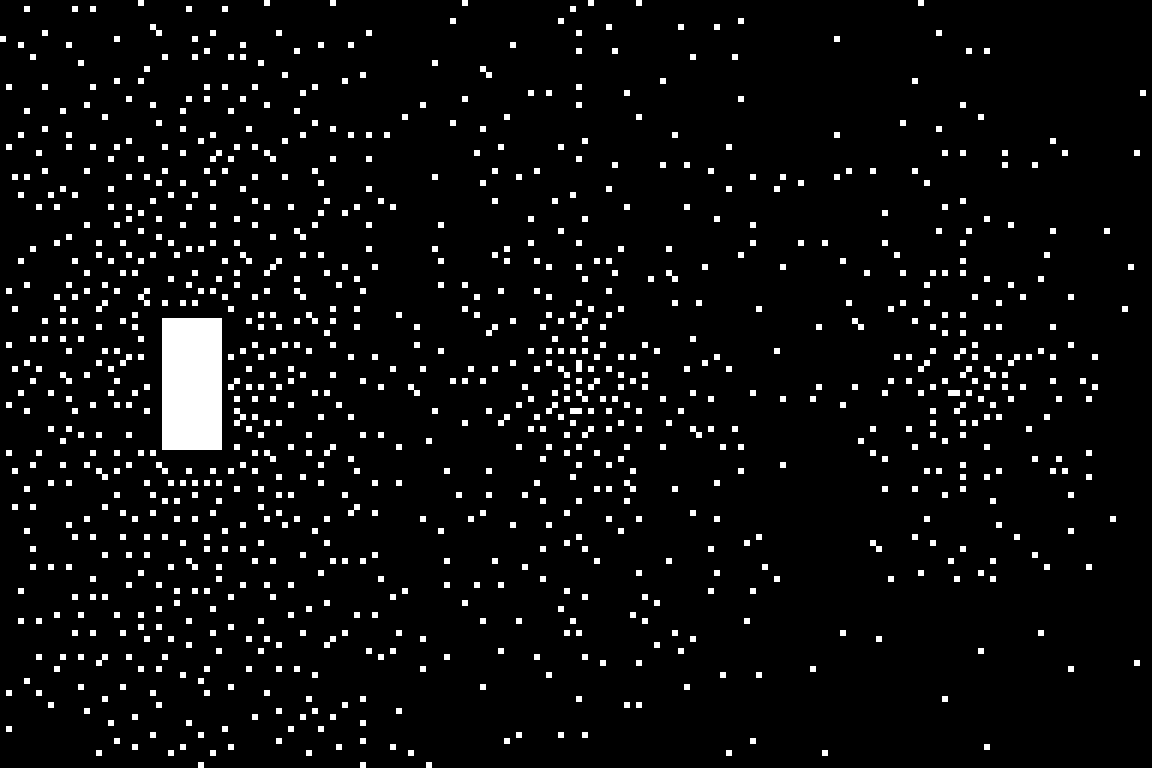}}%
\end{minipage}\,

\caption{Three frames for different relaxation times of the knee dataset, when
AF=24 was used, reconstructed with CS-DIC (a)-(b) and with CS-LR (e)-(f):
compare these with the corresponding fully sampled (FS) data in (g).
Poisson disk (c) and BASS optimized SPs (d) and (h) are also shown.}
\label{fig:fig-7}
\end{figure}

In Figure \ref{fig:fig-7}, visual results with the \textit{knee}
dataset illustrate the improvement due to using an optimized SP as
compared to using the Poisson disk SP, for both CS-LR and CS-DIC.
We also see that the optimized SPs are different for the two reconstruction
methods. Note that both optimized k-t-space SPs have a different sampling
density over time (first, middle, and last time frames are shown),
being more densely sampled at the beginning of the relaxation process.
The auto-calibration region in the first frame.

\subsection{BASS with a different criterion:}

\begin{figure}
\centering

\begin{minipage}[t]{0.48\columnwidth}%
\subfloat[]{\includegraphics[width=1\columnwidth]{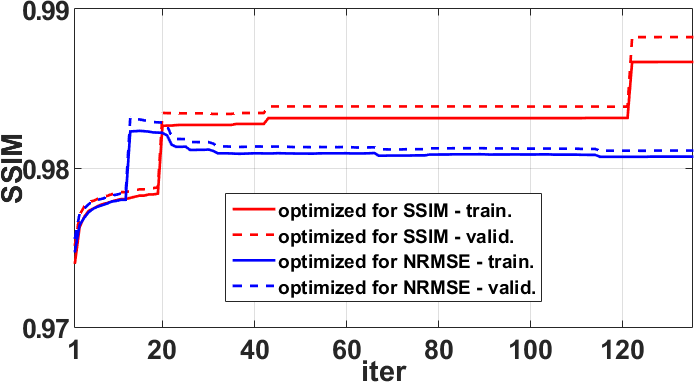}}%
\end{minipage}\,%
\begin{minipage}[t]{0.48\columnwidth}%
\subfloat[]{\includegraphics[width=1\columnwidth]{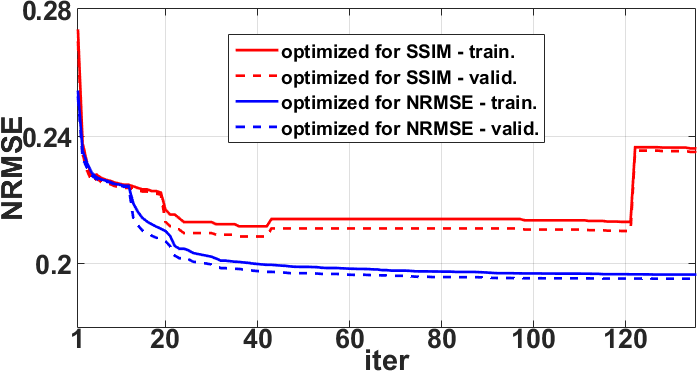}}%
\end{minipage}

\begin{minipage}[t]{0.32\columnwidth}%
\subfloat[]{\includegraphics[width=1\columnwidth]{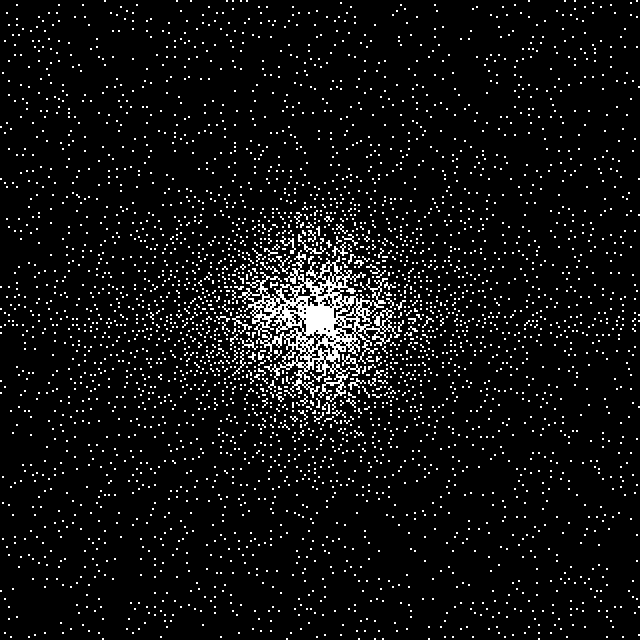}}%
\end{minipage}\,%
\begin{minipage}[t]{0.32\columnwidth}%
\subfloat[]{\includegraphics[width=1\columnwidth]{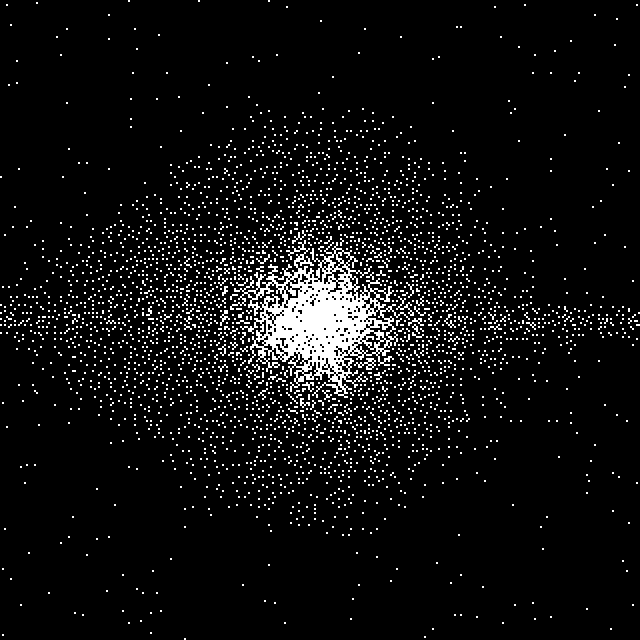}}%
\end{minipage}\,%
\begin{minipage}[t]{0.32\columnwidth}%
\subfloat[]{\includegraphics[width=1\columnwidth]{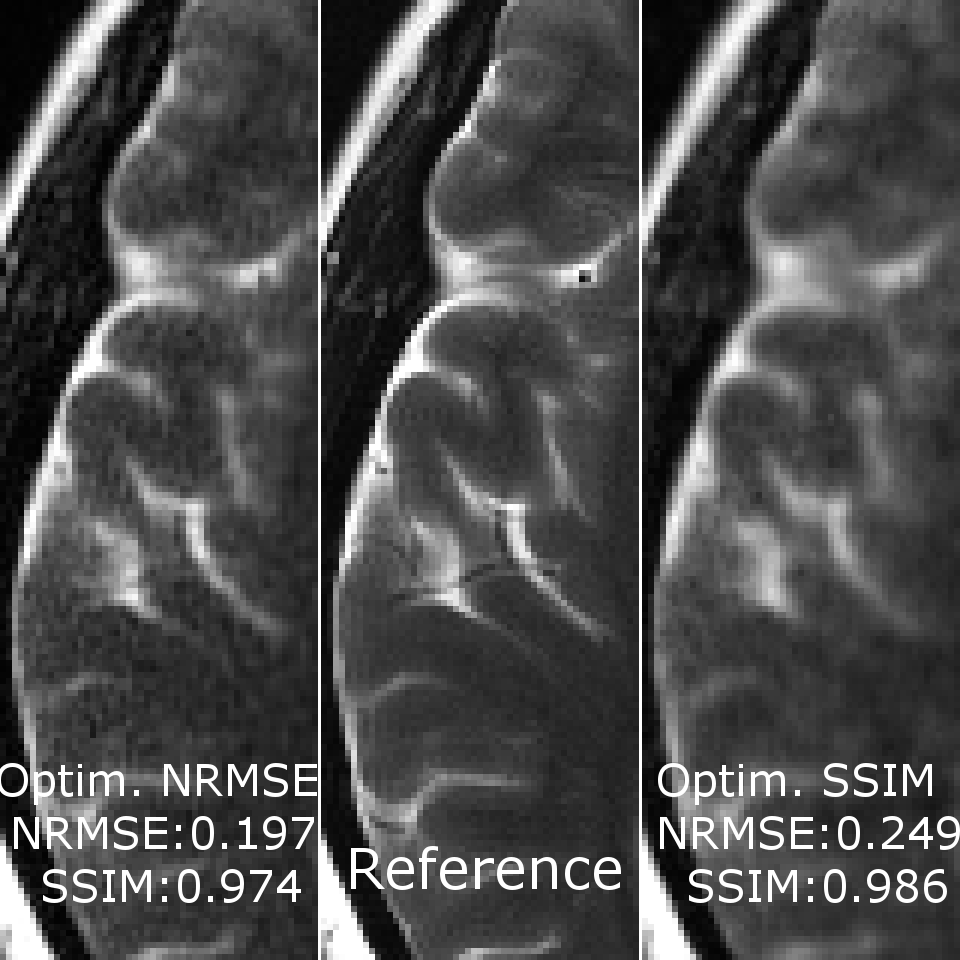}}%
\end{minipage}\caption{Comparing BASS in optimizing SSIM and NRMSE. (a) SSIM and (b) NRMSE
along the iterations, (c) SP obtained by optimizing SSIM, (d) SP obtained
by optimizing NRMSE, and (e) some visual results.}
\label{fig:fig-8}
\end{figure}

We illustrate that our proposed optimization approach is also efficacious
with different criteria. In some applications, one may desire the
best possible image quality, regardless of k-space measurements. Here
we discuss the use of BASS to optimize the SSIM of \cite{Wang2004a},
an image-domain criterion. For that, the task in (\ref{eq:DDO-general})
of finding the minimizer of $F(\Omega)$ in (\ref{eq:DDO-general-1}),
used in line \ref{if-F} of the Algorithm \ref{alg:proposed_algorithm},
is replaced by finding the minimizer in (\ref{eq:DDO-image-domain}),
with $g\left(\mathbf{x},\mathbf{y}\right)$ the negative of the SSIM.
In Figure \ref{fig:fig-8} we compare the optimization of SSIM with
that of NRMSE, using P-LORAKS on the \textit{brain} dataset, AF=16,
starting with the Poisson disk SP.

\section{Discussion\label{sec:DISCUSSION}}

The proposed approach delivers efficacious sampling patterns for high-resolution
or quantitative parallel MRI problems. Compared to previous approaches,
as in \cite{Gozcu2018,Gozcu2019,Sanchez2020,Duan-duanLiu2012,Ravishankar2011},
BASS is able to optimize much larger SPs, using larger datasets, spending
less computational time than greedy approaches (Figure \ref{fig:fig-2}a).
Greedy approaches test considerably more candidates SPs before updating
the SP. They are computationally affordable only for 1D undersampling
or small SPs.

The proposed approach is effective because it uses a smart selection
of new elements in the SP updating process. Candidates that are most
likely to reduce the cost function are tried first. The obtained efficient
solution may have minor differences depending on the initial SP (Figure
\ref{fig:fig-3}), but the optimized SPs tend to have the same final
quality if more iterations are used (Figure \ref{fig:fig-2}b). Adding
and removing multiple points at each iteration is beneficial for fast
convergence at the initial iterations (Figure \ref{fig:fig-2}c).

While the determination of the optimal number of training images is
still an open problem, we observed that using a large number of well-distributed
images in the training process helps to produce SPs that work better
with other unknown images (Figure \ref{fig:fig-2}d). The learned
SP can be used with new acquisitions of the same anatomy. This also
favors the use of fast training algorithms with larger datasets, instead
of computationally costly methods (such as greedy approaches) with
small datasets. 

The cost function in (\ref{eq:DDO-general-1}) evaluates the error
in k-space, not in the image domain. This may not be sufficiently
flexible because it does not allow the specification of regions of
interest in the image domain. Nevertheless, improvements measured
by the image-domain criteria NRMSE and SSIM were observed (Figure
\ref{fig:fig-5}). In different MRI applications other criteria than
(\ref{eq:DDO-general-1}) may be desired. The proposed algorithm can
be used for other criteria, such as the SSIM (Figure \ref{fig:fig-8}). 

The optimized SP varies with the reconstruction method (Figures \ref{fig:fig-6}
and \ref{fig:fig-7}) or with the optimization criterion (Figure \ref{fig:fig-8}):
thus sampling and reconstruction should be matched. This concept of
matched sampling-reconstruction indicates that comparing different
reconstruction methods with the same SP is not a fair approach, instead
each MRI reconstruction method should be compared using its best possible
SP. Note that the optimized SP improved the NRMSE by up to 45\% in
some cases (Figure \ref{fig:fig-5}).

The experiments also show that optimizing the SP is more important
at higher AFs. As seen in Figure \ref{fig:fig-5}, the optimization
of the SP flattened the curves of the error over AF, achieving a lower
error with the same AF. For example, P-LORAKS with optimized SP at
AF=20 obtained the same level of NRMSE as with variable density SP
at AF=6, while CS-LR with optimized SP at AF=30 obtained the same
level as with Poisson disk SP at AF=16, even after optimizing the
parameters used to generate the Poisson disk SP. These indicate that
it is possible to double the AF by optimizing the SP. Variable sampling
rate over time is advantageous for $\text{T}_{1\rho}$ mapping as
seen in \cite{Zhu2019}; it is very interesting that the algorithm
learned it, as shown in Figure \ref{fig:fig-7}.

Note that regularized algorithms, such as CS methods, may require
to estimate, or even optimize, the regularization parameter since
the best parameter is unknown a priori. Our approach is to optimize
the parameter using the initial SP, and re-optimize it with the learned
SP. However, joint optimization of the algorithm parameters and the
SP may be a better approach to be investigated in the future. This
is important for merging our approach with deep learning reconstruction,
where learning sampling and reconstruction happen simultaneously,
as recently in \cite{Bahadir2020a,Aggarwal2020a,Weiss2020}. That
appears to us likely to be the most suitable way for learned reconstruction
approaches \cite{Jacob2020,Knoll2019a,Liang2020,Wen2020}, as well
as for classical parametric inverse problems-based reconstructions
\cite{Fessler2020,Doneva2020,Sandino2020}. 

The lower computational cost and rapid convergence speed of BASS bring
the advantage of learning the optimal SP for various reconstruction
methods considering the same anatomy. Thus one can have a better decision
on which matched sampling and reconstruction is the most effective
for specific anatomy and contrast at the desired AF. Many questions
regarding the best way to sample in accelerated MRI can be answered
with the help of machine learning algorithms such as BASS. This new
learning method is expected to be a key element in making higher AFs
available in clinical scanners for translational research.

\section{Conclusions\label{sec:conclusions}}

We proposed a data-driven approach for learning the sampling pattern
in parallel MRI. It has a low computational cost and converges quickly,
enabling the use of large datasets to optimize large sampling patterns,
which is important for high-resolution Cartesian 3D-MRI and quantitative
and dynamic MRI applications. The approach considers data for specific
anatomy and assumes a specific reconstruction method. Our experiments
show that the optimized SPs are different for different reconstruction
methods, suggesting that matching the sampling to the reconstruction
method is important. The approach improves the acceleration factor
and helps with finding the best SP for reconstruction methods in various
applications of parallel MRI.

\section*{Acknowledgments}

We thank Azadeh Sharafi of CAI\textsuperscript{2}R for providing
knee MRI data and the fastMRI team for providing brain MRI data; also
Justin Haldar for fruitful discussions and for providing the P-LORAKS
codes and Mohak Patel for his codes on Poisson disk sampling.

\bibliography{library-10-29-2020}

\end{document}